\documentstyle[aps, preprint, epsf, axodraw]{revtex}
\tightenlines
\begin{document}

\newcommand{\be}{\begin{equation}}
\newcommand{\ee}{\end{equation}}
\newcommand{\bi}{\bibitem}
\newcommand{\bea}{\begin{eqnarray}}
\newcommand{\eea}{\end{eqnarray}}
\newcommand{\nn}{\nonumber}
\newcommand{\ps}{p\kern-0.175cm /}
\newcommand{\psm}{\left(\ps + m \right)}
\newcommand{\ks}{k\kern-0.19cm /}
\newcommand{\prk}{\frac{i}{\ps-\ks-m}}
\newcommand{\prkd}{\frac{i}{-2 p k  + k^2}}
\newcommand{\prkn}{\left( \ps - \ks + m \right)}
\newcommand{\prko}{\frac{i}{\ps-\ks_1-m}}
\newcommand{\prkon}{\left( \ps - \ks_1 + m \right)}
\newcommand{\prkod}{\frac{i}{-2 p k_1  +k_1^2}}
\renewcommand{\prc}{\frac{i}{\ps-\ks_c-m}}
\newcommand{\prbc}{\frac{i}{\ps-\ks_b-\ks_c-m}}
\newcommand{\prdbc}{\frac{i}{-2 p \left(k_b +k_c \right) + \left( k_b + k_c
\right)^2}}
\newcommand{\prdc}{\frac{i}{-2 p k_c + k_c^2}}
\newcommand{\prkko}{\frac{i}{\ps-\ks-\ks_1-m}}
\newcommand{\prkkon}{\left( \ps - \ks- \ks_1 + m \right)}
\newcommand{\prkkod}{\frac{i}{-2 p \left(k + k_1 \right) + 
\left( k + k_1 \right)^2}}
\newcommand{\gk}{\frac{-i}{k^2}}
\newcommand{\gko}{\frac{-i}{k_1^2}}
\newcommand{\gkc}{\frac{-i}{k_c^2}}
\newcommand{\gkb}{\frac{-i}{k_b^2}}
\newcommand{\gkbc}{\frac{-i}{(k_b+k_c)^2}}
\newcommand{\ph}{d^4 k d^4 k_1}
\newcommand{\phbc}{d^4 k_b d^4 k_c}

\newcommand{\gua}{\gamma^{\alpha}}
\newcommand{\guc}{\gamma^{\delta}}
\newcommand{\guai}{i g \gamma^{\alpha}}
\newcommand{\guat}{i g \gamma^{\alpha} T^a}
\newcommand{\guct}{i g \gamma^{\delta} T^c}
\newcommand{\gub}{\gamma^{\beta}}
\newcommand{\gubi}{i g \gamma^{\beta}}
\newcommand{\gubt}{i g \gamma^{\beta} T^b}
\newcommand{\gdai}{i g \gamma_{\alpha}}
\newcommand{\gda}{\gamma_{\alpha}}
\newcommand{\gdat}{i g \gamma_{\alpha} T^a}
\newcommand{\gdb}{\gamma_{\beta}}
\newcommand{\gdbi}{i g \gamma_{\beta}}
\newcommand{\gdbt}{i g \gamma_{\beta} T^b}
\newcommand{\go}{\gamma^{0}}

\newcommand{\sv}{S^{abc}_{\alpha \beta \delta}}
\newcommand{\cf}{C_F^2}
\newcommand{\ca}{\frac{C_F C_A}{2}}
\newcommand{\cfa}{\left(C_F^2-\frac{C_F C_A}{2}\right)}

\newcommand{\ub}{\bar{u}}
\newcommand{\mb}{\bar{m}}
\newcommand{\vea}{\left[vertex\right]_{\alpha}^a}
\newcommand{\veb}{\left[vertex\right]_{\beta}^b}

\newcommand{\fif}{C 5im_b^4}
\newcommand{\tif}{C 10m_b^3}
\newcommand{\zet}{\left( Z^{-1} - 1 \right)}

\begin{flushright}
UM-TH-98-08\\
hep-ph/9804401\\
\end{flushright}

\begin{center}
\vglue .06in
{\Large \bf {Cancellation of $1/m_Q$ Corrections to the Inclusive Decay Width of a
Heavy Quark.}}\\ [.5in]

{\bf A. Sinkovics, R. Akhoury and V. I. Zakharov }\\
[.15in]

{\it The Randall Laboratory of Physics\\
University of Michigan\\
Ann Arbor, MI 48109-1120}\\
[.15in]

\end{center}


\begin{abstract}
\begin{quotation}
We consider the infrared sensitivity of the inclusive heavy quark decay width
in perturbation theory. It is shown by explicit calculations to the second loop
order (when the non-abelian nature of the QCD interactions first become apparent)
that all infrared sensitive terms to logarithmic and linear accuracy cancel when the
decay width is expressed in terms of the short distance mass. This in turn implies
the absense of any non perturbative $1/m_Q$ power corrections to the decay width. The
cancellation is shown algebraically at the level of the Feynman integrands without
any use of an explicit infrared cutoff. This result is in accord with the
implications of the KLN theorem.
\end{quotation}
\end{abstract}
\newpage

\section{Introduction} \label{sec:intro}

The theory of heavy quark decays has attracted much interest lately in
connection with the decays of B mesons containing a single b-quark. Of 
particular phenomenological interest are the inclusive decays which can
provide information about the CKM parameters. Inclusive 
decays of hadrons containing a single heavy quark (mass $m_b$) have been
extensively studied \cite{hqope} within the operator product expansion (OPE) which
gives an expansion of the decay width in terms of a small parameter $\Lambda/m_b$ :
\be
\Gamma_{incl}~=~\Gamma^{pert}_{incl}(1+ a_1 \Lambda/m_b + a_2 ( \Lambda/m_b)^2
+...).
\label{rate}
\ee
In the above, $\Gamma^{pert}_{incl}$ is the decay width of the heavy quark
including all perturbative corrections (the parton model result). Since there are
no local gauge invariant operators of the appropriate dimension, an important
conclusion \cite{hqope} is that $a_1=0$, and the higher order power corrections
can be classified in terms of the matrix elements of various operators.

A great advantage of the OPE is the generality of its applicability to 
non-perturbative contributions as well. On the other hand, the conditions for
the validity of the OPE itself can be enunciated by explicit perturbative
calculations. As far as the power corrections are concerned, the method of
renormalons (for a review and further references see \cite{renorm}) may be used to
study these within perturbation theory. The basic idea of the renormalon method is
that one may obtain information about power corrections by looking at a class of
diagrams which give factorial divergence in large orders of perturbation theory.
The power corrections are seen to arise from regions of low momenta and thus we
are led to investigate the infrared sensitivity of inclusive decays. To this end
$\Lambda$, which is a non-perturbative parameter is replaced by an infrared cutoff
defined within perturbation theory. We will generically denote this infrared
cutoff by
$\lambda$ and in the case of QED, for example, it could be a (fictitious) photon
mass. Then the inclusive rate (\ref{rate}) may be represented as :
\be
\Gamma_{incl}~=~\Gamma^{pert}_{incl}(1+ b_1 \lambda/m_b + b_2 ( \lambda/m_b)^2 ln
\lambda  +...). \label{prate}
\ee
Note that the Landau conditions for the singularities of Feynman diagrams tell
us that the infrared sensitive contributions arise only as terms which are 
non-analytic in $\lambda^2$. Explicit calculations \cite{lincan} to leading order in
$\alpha$  have brought to light a subtlety that the coefficient $b_1$ is zero in
perturbation theory only if the decay rate on the right hand side of 
Eq. (\ref{prate}) is expressed not in terms of the pole mass $m_b$ but rather in
terms of a short distance or a $\bar{MS}$ mass. Thus perturbation theory is in
full accord with the OPE, only if in (\ref{rate}) the short distance mass is
used. In this way one does not have to worry about an otherwise large (possibly
non-universal) correction of order $1/m_b$, but only at the expense of
introducing a renormalization scale ($\mu$) dependence in the total rate through
the use of the running mass. We will discuss below that this result arises from
the cancellation of the infrared sensitive contributions proportional to
$\lambda /m_b$ (and of course to $ ln \lambda$) arising from the long distance
pieces of the mass and wave function renormalizations of the initial heavy quark
and from  the gluon radiation.
 
The result mentioned at the end of the previous paragraph has a simple
physical interpretation. The decay rate may have two possible sources
\cite{fnote1} of infrared sensitivity : (a) the heavy quark mass and (b) the
energy radiated during the decay. To expose the infrared sensitivity of the
heavy quark mass say within QED, consider a $Q\bar{Q}$ pair separated by a
distance
$r$ and interacting by the exchange of photons of mass $\lambda$. The total
energy is
\be
E~=~2m_b - {\alpha_{el} \over r}exp(-\lambda r) \sim 2m_b( 1+{ \alpha_{el} \lambda
\over 2 m_b }) - \alpha_{el}/r
\ee
where we have kept only the term linear in $\lambda$. It is clear from this
that we may identify
\be 
 m_b^{IR}~=~ \alpha_{el} \lambda /2,
\ee
 with the infrared sensitive
piece of the heavy quark mass and for small $r$ it is appropriate to define the
short distance mass $\bar{m}$ by:
\be
\bar{m}~=~m_b(1+ { \alpha_{el} \lambda \over 2 m_b })
\ee
The same result was obtained in \cite{lincan} by the usual renormalon method . 
Let us next consider the radiated energy. If the decay time is $\tau$ then 
classically, a light signal cannot reach a distance beyond
\be
R_{causal}~=~c \cdot \tau
\ee
As a result, the field energy stored at distances larger than $R_{causal}$
cannot be transformed into the energy of the decay products and cannot
affect the decay rate and hence must be radiated away. Let us denote this long
distance piece of the energy radiated during the decay by $E_{rad}^{IR}$.
That this long distance part of the radiated energy is proportional linearly to
the infrared cutoff may be seen from the classical expression for it:
\be
E_{rad}^{IR}~=~\int_{small~ \omega}d \omega I(\omega).
\ee
$ I(\omega)$ is the intensity of radiation which approaches a constant
\cite{jackson} as
$\omega \rightarrow 0$. The cancellation of the linear term in 
Eq. (\ref{prate}) is the statement that to this order,
\be
m_b^{IR}~=~E_{rad}^{IR}
\ee
and the decay rate which is sensitive to the energy release at short distances
is not affected.

Certain questions immediately come to mind concerning the generality of these
results. Is there a general principle behind this cancellation ?, and is it true
to all orders in perturbation theory, even for the non-abelian case? In fact
there are infrared safe observables in QCD that do receive power corrections
that are proportional to $1/M$ where $M$ is a large mass scale. Examples are
provided by the event shape variables in $e^+e^-$ annihilation \cite{thrust} for
which there does not exist an operator product expansion. It has been argued in
ref. \cite{az1} that inclusive enough observables do not receive power
corrections of the type
$1/M$ whereas those that assume some precision measurement (and hence are more
exclusive) like the event shapes, in fact do. Below we will apply this general
principle to the inclusive decay of the heavy quark. We will argue that the KLN
 \cite{kln} theorem is behind the cancellation of the terms
proportional to $1/m_b$. The purpose of this paper is to show by an  explicit
calculation in the full QCD theory the cancellation of the infrared sensitive
pieces that are linear in the infrared cutoff to order $\alpha_s^2$ in the decay
rate. 

In a recent publication \cite{asz} it was argued that when the transition
probability is   summed over both initial and final degenerate states as is
envisaged in the KLN theorem, not only the leading infrared sensitive piece
proportional to $ln \lambda$ but the one proportional to $\lambda$ is
cancelled as well, i.e,
\be
\sum_{i,f} |S_{i\rightarrow f}|^2 \sim 0 \cdot ln \lambda + 0 \cdot 
\lambda + {\rm terms~independent~of~ \lambda} + O(\lambda^2 ln \lambda).
\ee
The transition probability summed over both initial and final states cannot be
related to a physical inclusive cross section. However we will now argue that
for heavy quark decay, the initial state is trivial and hence, the above
mentioned property of the KLN theorem guarantees the cancellation not only of
the $\ln \lambda$ terms but also those proportional to $\lambda$ itself.
To substantiate this we note that the KLN theorem presumes that the total
energy can be fixed to any accuracy. This is in the spirit of the uncertainity
principle, which allows us to measure the total energy to any accuracy provided
the measurement time is long enough.  In particular, the uncertainity $\Delta
E$ in the total energy can be made smaller than an infrared cutoff:
\be
\Delta E \ll \lambda
\ee
Now consider a decaying charged (or colored) particle in its rest frame. It is
obvious that this state is not degenerate with any other since there is a
gap between the mass of the charged particle and the energies in the continuum
for any $\lambda \neq 0$. This is most clearly seen for QED with a non-zero
photon mass, but is true for any infrared cutoff on the energy, $\lambda$. 
Thus 
the summation over the initial states in the KLN theorem is redundant
for this case. The final states however include the full degeneracy but the
summation over the final states is equivalent to calculating the decay width, and
hence we conclude that,
\be
\Gamma_{incl}~=~\Gamma^{pert}_{incl}\left(1 + O(\lambda^2 ln \lambda)\right).
\ee
It should be emphasized that this argument guarantees the above equality to all
orders in perturbation theory. However we should note that it is very important
for the KLN theorem to be valid that the renormalization procedure does not
introduce any infrared sensitivity. Thus since the pole mass is infrared
sensitive, on shell mass renormalization must be avoided in favour of the
$\bar{MS}$ scheme . In concluding this we would like to point out that the same
argument can be applied not just to the semileptonic decay but to other inclusive
decays like for example the radiative one, $B \rightarrow X_s \gamma$. In fact,
the explicit perturbative calculation though performed for the semileptonic
decay, can equally well be applicable for the radiative decay $b \rightarrow s
\gamma$ with minor changes for the electroweak vertices and phase space factors.

As mentioned previously, we will explicitly verify the above argument in this
paper for the inclusive semileptonic decay of a heavy quark in QCD to the
second loop order when the non-abelian nature of the interactions are first
apparent. That is we show the cancellations of all terms linear in the infrared
cutoff in the decay width when the latter is expressed in terms of the short
distance mass. The possible sources of infrared
sensitive terms  are the mass and wave function
renormalization diagrams of the initial heavy quark and the
bremsstrahlung diagrams. We will discuss the cancellation of the logarithmic and
the linearly infrared sensitive terms in the diagrams for the wave function
renormalization and of the linearly IR sensitive terms in those for the mass
renormalization. These cancellations proceed in a different manner for the two
cases because whereas the wave function renormalization is multiplicative, the
on-mass shell renormalization is additive. In section \ref{sec:prelim} 
we introduce our notations and conventions and discuss our strategy. 
In particular we show how to the linear
accuracy desired, we may replace the diagrams for the radiation from the final
state massless quark by effective local vertices. Next we discuss the wave
function renormalization in 
section \ref{sec:wave} and the mass renormalization in section \ref{sec:mass}.
Certain additional technical details are relegated to the appendix. 
In section \ref{sec:summary} we summarize our results and discuss future 
prospects.

\section{Preliminaries} \label{sec:prelim}
We consider the totally inclusive semileptonic decay 
of a heavy quark and examine the possible term linearly proportional to an
infrared cutoff. Such a term can arise from the soft gluon contributions to the
diagrams for the mass and wave function renormalizations of the heavy quark and
from the bremsstrahlung diagrams.  

The decay rate is proportional to the imaginary part
of the forward amplitude shown in Fig. \ref{fig:fsa}. 
The hard amplitude involving the final
state quark is denoted by the shaded blob which also includes the lepton loop. The
soft gluon interactions responsible for the infrared sensitivity dress the hard
amplitude, connecting to it and to the heavy quark. We will use the
fact that for the soft gluon of momenta $k
\ll m_b$, we may perform the sum over cuts implicit in taking the imaginary part,
for just the hard part of the amplitude leaving the soft gluon lines uncut. This
simplification arises due to the fact that for the process of a heavy quark decay,
we may  assign a positive energy flow direction to each soft gluon consistently.
Then we may  write for the corresponding gluon propagator:
\be
{i \over k^2+i\epsilon} ~= ~2\pi\theta(k_0)\delta(k^2)+{i \over k^2+ik_0\epsilon}
\ee
The second term can never cause a pinch singular point since the contour can be
deformed into the lower half $k_0$ plane. This piece therefore (from the Landau
equations) does not give any contributions that are non-analytic in the infrared
cutoff. Thus as far as the infrared sensitive terms are concerned, even upto
linear accuracy it does not matter if we cut the gluon propagator or not, both
possibilities giving the same contribution. Keeping this in mind, we will sum over
the cuts of the hard amplitude only and leave the soft gluon propagators uncut in
the diagrams for the forward scattering amplitude. The next simplification of the
forward scattering amplitude we will use for our infrared analysis is that the
(heavy quark) propagators on the right hand side of the cut can also be taken to
have the
$+i\epsilon$ prescription, the same as those on the left hand side of the cut.
The difference between the two choices is readily seen to be pure imaginary and
since the cut hard amplitude is already imaginary, we may ignore this
difference for our purposes.

Let us consider the diagrams with the soft gluon radiation interacting with the
hard part.  In the next subsection we will show that to the accuracy we want,
after summing over the cuts of the hard part, and integrating over its phase
space and that of the leptons, this interaction may be replaced by effective
local vertices for the interaction of one and two gluons. These effective
vertices may be obtained from the corresponding one without gluon radiation by
gauge invariance. Thus we can forget about the details of the hard part and use
the effective vertices to discuss the infrared sensitive contributions for such
diagrams. Diagrams in which all the gluon interactions are contained in the hard
part are irrelevant to our analysis as far as the infrared sensitive terms are
concerned which are linearly proportional to the infrared cutoff. Since the wave
function renormalization is multiplicative, all its contributions factorize from
the hard amplitude. In addition, as we discuss below the corresponding
bremsstrahlung diagrams that cancel the logarithmic and linear dependence on the
IR cutoff of the wave function renormalization diagrams may be obtained by
certain Ward-like identities. These cancellations can be shown independent of 
the
hard amplitude whose precise structure is not needed here, but the latter is
essential for showing the cancellations among the mass renomalization and the
corresponding radiation type diagrams. The use of the heavy quark rest frame
provides very important simplifications of this analysis.

In this paper all the infrared cancellations are shown strictly algebraically    
at the level of the corresponding  Feynman integrands. Power counting is used to
isolate and to show the cancellations of the possibly infrared sensitive
contributions upto linear accuracy. After the cancellations, from power counting,
the integrand is seen to contribute something which would at least be 
proportional quadratically to the infrared cutoff. In this way we avoid using any
particular IR cutoff. This is crucial since for the non-abelian theory we do not
know of any gauge invariant infrared cutoff for explicit calculations at the two
loop level.

Finally we note that throughout this paper we work in the Feynman gauge. The
cancellation of the infrared sensitive terms is a gauge invariant statement,
however, the type of diagrams contributing to this cancellation depends on the
choice of gauge.

\subsection{Effective vertices for the Interaction of radiation with
the Hard Amplitude} 
In this section we will derive the effective vertices for the interaction of
the gluon radiation with the final state massless quark in the hard amplitude..
These effective local vertices are constructed to reproduce the soft
momentum ($k \ll  m_b$) contributions upto linear in $1/m_b$ accuracy. Consider
the absorbtion of a single gluon of momentum $k$ from the final state quark. We
will show below that summing over the cuts for the final state fermion line
cancels the term proportional to $1/k$ (i.e., the would be infrared divergent
term proportional to
$ln\lambda$ ) so that the leading left over piece is of order $k^0$. When the
other end of the gluon line is attached to the heavy quark line then we would be
interested in the term proportional to $\lambda$ and from simple power counting of
the corresponding diagram we see that we need only keep this leading term
proportional to $k^0$ from the sum over cuts of the final state quark. In this
way we will conclude that after the phase space integration over the leptons and
the final state quark at fixed $k$ this effective interaction may be represented 
by a local vertex of the form shown in Fig. \ref{fig:effvertex}b). 
We show below that this effective
local vertex is what we would obtain by the substitution  
$p \rightarrow (p-gA)$ 
in the leading effective vertex of Fig. \ref{fig:effvertex}a). 
A similar effective vertex is derived for the
absorbtion of two gluons from the final state quark. Since as mentioned earlier,
the heavy quark wave function renormalization contribution to the infrared
sensitivity is multiplicative in nature, these contributions are not 
sensitive to
the details of the interaction of the final state quark and the 
effective vertices
derived in this section are not needed there.

In the parton model, the total width for the semileptonic decay is to lowest
order: 
\be
\Gamma_0~=~{ G_F^2|V_{ub}|^2 \over 192 \pi^3}m_b^5
\ee
We may represent this by an effective local vertex, obtained by
preforming the phase space integration over the leptons and and the massless
quark by the heavy dot in Fig. \ref{fig:effvertex}a). 
Such an effective vertex may be written as:
\be
C\bar{u}\ps^5 u \label{v0}
\ee
where, $u$ is the heavy quark spinor.  In the above, 
\be
C~=~{ G_F^2|V_{ub}|^2 \over 192 \pi^3}.
\ee
The vertex of Fig. \ref{fig:effvertex}a) 
will be referred to as the leading effective vertex.
Next consider the class of diagrams of Fig. \ref{fig:cut}a). 
The gluon may undergo self
interactions before connecting to the heavy quark line. Thus, for example, the
diagrams of Fig. \ref{fig:cut}b) are included in the set. 
Next we sum over the two cuts on the
massless quark line. In one just the quark line is cut and in the other, both
the quark and gluon lines are cut. However, as discussed above and as it 
was shown in
explicit examples in ref. \cite{asz}, as far as the infrared sensitive terms
(those  that are non-analytic in the infrared cutoff) are concerned, both 
real and virtual soft quanta may be regarded the same i.e, 
we may write for the gluon line
contribution either as $2\pi\theta(k_0)\delta(k^2)$ or
$i/(k^2+i\epsilon)$. This is really  a consequence of the Landau equations as
discussed earlier. Keeping this in mind we can write for the sum over cuts in 
Fig. \ref{fig:cut}a) as:
\be
-\Gamma^{*}\ps_1\gamma^{\alpha}gT^a \left(\ps_1-\ks \right)
\Gamma 2\pi \left[{\delta(p_1^2) \over (p_1-k)^2+i\epsilon} + 
{\delta((p_1-k)^2) \over
p_1^2-i\epsilon} \right], \label{cut1}
\ee
where $\Gamma$ generically denote the weak interaction vertices, and
$p_1=(p-q)$ with $q$ the momentum transfer to the leptons. Note that the lepton
loop is always cut and we do not explicitly write down the corresponding
expression. We also have not explicitly written down the contributions from the
blob in Fig. \ref{fig:cut}a) and those from the external quark lines since 
their exact form is inessential to the argument. 
Now we can use the following identity to simplify the term in squared 
brackets above:
\bea
{\delta(p_1^2) \over (p_1-k)^2+i\epsilon} + {\delta((p_1-k)^2) \over
p_1^2-i\epsilon} ~=~ \nonumber \\
(-1)\int_0^1dx_1\int_0^1dx_2\delta(1-x_1-x_2)\delta^{(1)}
\left(x_1p_1^2+x_2(p_1-k)^2 \right).
\label{id1}
\eea
In the limit $k\rightarrow 0$ the right hand side of Eq. (\ref{id1}) becomes:
\be
(-1) \delta^{(1)}(p_1^2),
\ee
which is a well defined quantity. This shows the cancellation of the leading
infrared divergent term, and the next term in the sum over cuts, 
which is $O(k^0)$ may be written as:
\be
-\Gamma^{*}\ps_1\gamma^{\alpha}gT^a\ps_1
\Gamma 2\pi (-1) \delta^{(1)}(p_1^2) \label{cut12}
\ee
We wish to compare this with the corresponding expression for the absorbtion of a
zero momentum gluon from the uncut final state massless quark of momentum
$p_1$ which is:
\be
i\Gamma^{*}\ps_1\gamma^{\alpha}gT^a\ps_1
\Gamma {1 \over (p_1^2+i\epsilon)^2}.
\ee
Using the fact that:
\be
2i Im {1 \over (p_1^2+i\epsilon)^2} ~=~2\pi\delta^{(1)}(p_1^2),
\ee
and comparing with Eq. (\ref{cut12}) we see that  as far as the terms of order
$k^0$ are concerned we may represent the diagram of Fig. \ref{fig:cut}a) 
for fixed $k$ as an
effective local interaction which after the appropriate phase space integrations
over the leptons and the massless quark for a fixed $k$ may be obtained from the
effective vertex Eq. (\ref{v0}) by means of the replacement $p_{\mu} \rightarrow
(p_{\mu}-gT^aA^a_{\mu})$. We have of course shown this only for the one gluon
effective vertex shown in Fig. \ref{fig:effvertex}b). 
Next we will show that the same substitution also reproduces the two gluon 
emission or absorbtion process. 

Consider next the absorbtion of two gluons by the final state quark as shown in
Fig. \ref{fig:cut}c) and once again the blob represents all possible 
interactions and connections of these gluons with each other and with the 
heavy quark. The sum over cuts of the massless quark line for this 
configuration for fixed $k_1$, $k_2$ is:
\bea
\Gamma^{*}\ps_1 gT^a\gamma^{\alpha}\left( \ps_1-\ks_2 \right)
gT^b\gamma^{\beta} \left(\ps_1-\ks_1-\ks_2 \right)\Gamma
2\pi\left[ \delta(p_1^2) {1 \over (p_1-k_2)^2+i\epsilon}{1 \over
(p_1-k_1-k_2)^2+i\epsilon} + \right. \nonumber \\
\left. \delta\left((p_1-k_2)^2\right) {1 \over p_1^2-i\epsilon}{1 \over
(p_1-k_1-k_2)^2+i\epsilon} + 
\delta\left((p_1-k_1-k_2)^2 \right) {1 \over p_1^2-i\epsilon}{1 \over
(p_1-k_2)^2-i\epsilon} \right]
\eea
Next we use a generalization of the identity Eq. (\ref{id1}), 
\bea
&& \delta(p_1^2) {1 \over (p_1-k_2)^2+i\epsilon}{1 \over
(p_1-k_1-k_2)^2+i\epsilon} + 
\delta\left((p_1-k_2)^2 \right) {1 \over p_1^2-i\epsilon}{1 \over
(p_1-k_1-k_2)^2+i\epsilon} + \nonumber \\
&& \delta\left((p_1-k_1-k_2)^2\right) {1 \over p_1^2-i\epsilon}{1 \over
(p_1-k_2)^2-i\epsilon} 
=(-1)^2\int_0^1..\int_0^1dx_1dx_2dx_3\delta\left(1-x_1-x_2-x_3 \right) 
\times \nonumber \\
 && \delta^{(2)}\left(x_1p_1^2+x_2(p_1-k_2)^2+x_3(p_1-k_1-k_2)^2 \right) 
\label{id2}
\eea
The right hand side of (\ref{id2}) tends to a well defined limit as $k_i
\rightarrow 0$,
\be
(-1)^2{1 \over 2} \delta^{(2)}(p_1^2)
\ee
which again represents the cancellation of the leading logarithmically divergent
contribution. The term proportional to $k^0$ in the sum over cuts is then given
by:
\be
\Gamma^{*}\ps_1gT^a\gamma^{\alpha}\ps_1
gT^b\gamma^{\beta}\ps_1\Gamma
(-1)^2{1 \over 2} \delta^{(2)}(p_1^2). \label{cut2}
\ee
Again consider the corresponding term for the absorbtion of two zero
momentum gluons from the massless quark line:
\be
-i\Gamma^{*}\ps_1gT^a\gamma^{\alpha}\ps_1
gT^b\gamma^{\beta}\ps_1\Gamma
{1 \over (p_1^2+i\epsilon)^3}
\ee
Since
\be
2i Im {1 \over (p_1^2+i\epsilon)^3} ~=~-\pi\delta^{(2)}(p_1^2),
\ee
comparing with (\ref{cut2}) we see that the $O(k^0)$ contribution from the 
sum over cuts may again be written as an effective local vertex after the 
relevant phase space integrations. This local vertex can be obtained from 
Eq. (\ref{v0}) by means of the replacement $p_{\mu}\rightarrow (p_{\mu}-
gT^aA^a_{\mu})$ and expanding to $O(g^2)$. Note that there is a similar 
contribution as in Eq.(\ref{cut2}) but with the interchange 
$(\alpha,a )\leftrightarrow (\beta,b)$, and this too is reproduced by the 
above mentioned replacement. The two gluon effective vertex is shown in 
Fig. \ref{fig:effvertex}c).

We should remark here that when we go beyond the two gluon emission case, in
general multi-gluon emission or absorbtion from the final state line cannot be
written as an effective local vertex. Fortunately this is not a problem for us.

\section{Cancellation of Infrared Sensitive Terms Upto Linear Accuracy from 
the Wave Function Renormalization Diagrams} \label{sec:wave}
In this section we will show that the infrared sensitive terms upto linear
accuracy are cancelled between the wave function renormalization contributions
from diagrams with self energy insertions and parts of other diagrams related  
to the bremsstrahlung. The cancelling sets can be arranged into groups which as we
will see below upto linear accuracy are obtained through certain Ward-like
identities.  For the cancellations considered in this section which are related
to the wave function renormalization, the corresponding contributions are such
that the hard part of the diagrams factorize. The hard part thus is not
consequential to the argument and just comes along for the ride.

Throughout the paper, we will denote by $-i\Sigma(p)$ the sum of all 2 point one
particle irreducible graphs. Then the full fermion propagator is:
\be
S_F^{'}(p)~=~{i \over \ps-m_b-\Sigma(p)}
\ee
Near the mass shell we have
\be
\Sigma(p)~=~\delta m\left(m_b,\Delta m \right) - (Z^{-1}-1)
\left(\ps-m_b \right) + O\left((\ps-m_b)^2 \right)
\ee
where $\Delta m$ denotes the mass counterterm:
\be
\delta m\left(m_b,\Delta m \right) \left. \right|_{\ps=m_b}~=~0.
\ee
Thus $m_b$ denotes the pole or the physical mass:
\be
\lim_{\ps \rightarrow m_b}S_F^{'}(p)~=~{i \over \ps-m_b}Z
\ee
From the LSZ reduction formula, each external line in the S-matrix element 
gets a factor $Z^{1/2}$ and we are interested in its perturbative expansion. 
Thus we expand,
\be
Z^{1/2}~=~1-{1 \over 2}\left(Z^{-1}-1 \right)+{3 \over 8}\left(Z^{-1}-1
\right)^2+..... \label{wf1}
\ee
and $ Z^{-1}-1 $ may be computed using:
\be
Z^{-1}-1~=~-{\bar{u}{\partial \over \partial p_{\nu}}\Sigma \left.
\right|_{\ps=m_b} u \over
\bar{u}\gamma^{\nu}u} \label{wf2}
\ee
In the rest frame of the heavy quark, only $\nu=0$ contributes.

We will now clarify our procedure first with the $O(\alpha)$ case and next the
two loop example will be discussed.

\subsection{Cancellation to One-Loop Order}

Consider the graphs of Fig. \ref{fig:wave1}. 
The wave function renormalization parts of both
graph(a) and graph(b) are the hard part times a factor of $Z^{1/2}$. Thus we 
expand it using (\ref{wf1}) keeping only the first two terms and computing the
perturbative contribution to order $\alpha$ of $Z^{-1}-1$ from (\ref{wf2}).
Next using the identity
\be
{\partial \over \partial p^{\nu}}{i \over \ps-\ks-m_b}~=~
{i \over \ps-\ks-m_b}i\gamma_{\nu}{i \over
\ps-\ks-m_b}
\ee
we obtain the Ward-like identity  shown in Fig. \ref{fig:ward1}, 
where the dotted line just denotes the insertion of a zero momentum vertex 
$\gamma_{\nu}$. We would now like to show that upto logarithmic and linear 
accuracy in the infrared, the right
hand side of Fig. \ref{fig:ward1} may be identified with the wave function 
renormalization part of Fig. 
\ref{fig:wave1}c) times $\bar{u}\gamma_{\nu}u$, 
modulo the hard part, which after the relevant phase space integrations just 
gives a factor of $C m_b^5$. It is easiest to go now to the rest frame of the
heavy quark, where
\be
\gamma^0 u~=~u. \label{hqu}
\ee
From
this and Eqns. (\ref{wf1}), (\ref{wf2}) we immediately see 
\be
\left. \ref{fig:wave1}c) \right|_{wave function}= (Z^{-1}-1)_{g^2} Cm_b^5 \ub u
= -\left[ \ref{fig:wave1}a) + 
\ref{fig:wave1}b) \right].
\ee
Here, and henceforth, unless otherwise specified, the equality sign is for terms
that are logarithmically and linearly infrared sensitive. This is the desired
result, which does not involve using an explicit infrared cutoff. 
Let us now see how the wave function renormalization part of 
Fig. \ref{fig:wave1}c) can be identified with 
$(Z^{-1}-1)_{g^2} Cm_b^5$, which is effectively the right hand side of Fig. 
\ref{fig:ward1} multiplied by the hard part $Cm_b^5$.
Consider the expressions for
these diagrams in the rest frame of the heavy quark. Both are of the form:
\be
(ig)^2 C_F C m_b^5 \int {d^4k \over (2\pi)^4} N {i \over -2pk+k^2}{i
\over-2pk+k^2}{-i \over k^2}
\ee
where $N$ denotes the numerators $N_{5}$ or $N_{4c}$ for the two diagrams
respectively. Thus we only have to compare the numerators to linear accuracy. 
Upto linear terms in the infrared power counting
\bea
N_{5}=&&\bar{u}\gamma^{\alpha}(\ps+m)\gamma^0
(\ps+m)\gamma_{\alpha}u -
\bar{u}\gamma^{\alpha}\ks\gamma^0(\ps+m)\gamma_{\alpha}u \\
 \nonumber
&&-\bar{u}\gamma^{\alpha}(\ps+m)\gamma^0\ks\gamma_{\alpha}u
\eea
There is a similar expression for the numerator $N_{4c}$ 
except for the factor of $\gamma_0$. However, using 
Eq. (\ref{hqu}) it is easy to see that to this accuracy the two numerators 
are equal 
\be
N_{5}=N_{4c}
\ee
This demonstrates the cancellation of not only the infrared divergent part but
also the next l inearly infrared sensitive piece in the wave function
renormalization.
After the cancellation, the integral will be at most quadratically divergent
by  IR power counting.
The cancellation of the wave function parts at two loops follows the same 
general method except that now we have more diagrams.

\subsection{Cancellation to Two Loop Order} \label{sec:wave22} 
The 2-loop one particle irreducible diagrams can be divided
into  5 groups according to the 5 pieces of the order $g^4$  self-energy
corrections  (Fig. \ref{fig:selfen}).
These groups are essentially the different color groups, 
with the exception that the first and third group have the same color structure.
In this section we 
show the cancellation of infrared sensitive terms to linear accuracy in the
second loop order for the wave function renormalization diagrams and their
radiation counterparts.  Power counting in
the IR region is used to isolate logarithmically and linearly 
infrared sensitive pieces of the integrals. In the following, these
infrared sensitive terms identified by power counting will be simply 
referred as ``logarithmic and linear pieces'' of the corresponding diagrams. 
As it is discussed
below, the cancellation of logarithmic and linear terms takes place 
within each color group separately. 
The cancellation is algebraic at the level of integrands
and no explicit infrared cutoff is used.

\begin{figure}
\begin{center} \begin{picture}(400,90)(0,0)

\SetScale{7}
\SetWidth{0.071}

\SetOffset(0,12)
\Line(1,1)(5,10)
\Line(9,1)(5,10)
\Vertex(5,10){0.5}
\GlueArc(3,5.5)(1.7,66,246){-0.23}{9}
\GlueArc(2.3,3.94)(1.7,-114,66){-0.23}{9}
\Text(35,0)[]{a)}

\SetOffset(84,12)
\Line(1,1)(5,10)
\Line(9,1)(5,10)
\Vertex(5,10){0.5}
\GlueArc(7,5.5)(1.7,-66,114){-0.23}{9}
\GlueArc(7.69,3.94)(1.7,114,294){-0.23}{9}
\Text(35,0)[]{b)}

\SetOffset(168,12)
\Line(1,1)(5,10)
\Line(9,1)(5,10)
\Vertex(5,10){0.5}
\Gluon(2.55,4.5)(7.45,4.5){-0.23}{9}
\GlueArc(2.655,4.73)(1.7,66,246){-0.23}{9}
\Text(35,0)[]{c)}

\SetOffset(336,12)
\Line(1,1)(5,10)
\Line(9,1)(5,10)
\Vertex(5,10){0.5}
\Gluon(2.55,4.5)(7.45,4.5){-0.23}{9}
\GlueArc(7.345,4.73)(1.7,-66,114){-0.23}{9}
\Text(35,0)[]{e)}

\SetOffset(252,12)
\Line(1,1)(5,10)
\Line(9,1)(5,10)
\Vertex(5,10){0.5}
\Gluon(3.88,7.5)(7.88,3.5){-0.23}{9}
\Gluon(2.12,3.5)(6.12,7.5){-0.23}{9}
\Text(35,0)[]{d)}

\end{picture}

\end{center}
\caption{\setlength{\baselineskip}{0.30in}
The second group of 2-loop diagrams corresponding to $\Sigma_{22}$.}
\label{fig:wave22}

\end{figure}
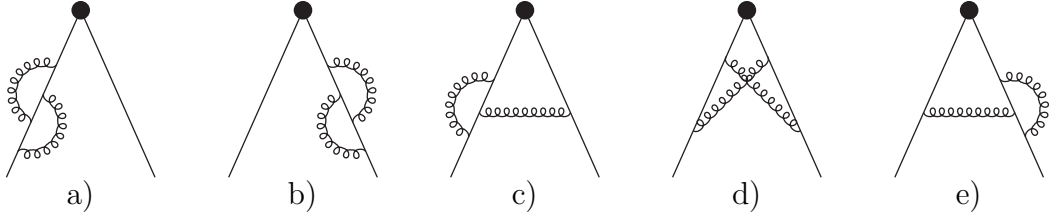

Let us take the second group (Fig. \ref{fig:wave22}) as an example 
to work with. We would like to show that both the logarithmic and the 
linear pieces in the wave function
renormalization of diagrams \ref{fig:wave22}a) and 
\ref{fig:wave22}b) cancel the corresponding
logarithmic and linear wave function renormalization pieces of diagrams 
\ref{fig:wave22}c), \ref{fig:wave22}d) and \ref{fig:wave22}e), i.e. 
\bea 
&& \left[ \ref{fig:wave22}a) + \ref{fig:wave22}b) \right]_{wave function}= 
-C m_b^5 \zet_{g^4, \Sigma_{22}} \ub u \\ \nn
&&
= -\left[ \ref{fig:wave22}c) + \ref{fig:wave22}d) +\ref{fig:wave22}e) \right]
_{wave function}
\eea
As before, we may write
\be
\zet_{g^4, \Sigma_{22}} = -\frac{\ub \left. 
\frac{\partial{\Sigma_{22}}}{\partial p_{\nu}} \right|_{\ps = m_b} 
u }{\ub \gamma^{\nu} u} =-\frac{\ub \left.
\frac{\partial{\Sigma_{22}}}{\partial p_{0}} \right|_{\ps = m_b}
u }{\ub u}
\ee
where the last equality holds in the rest frame of the heavy quark
$p=(m_b,0,0,0)$.

We would like to show that to linear accuracy
\bea 
&& C m_b^5 \ub \left. \frac{\partial{\Sigma_{22}}}{\partial p^{0}} 
\right|_{\ps = m_b}u  =
i C m_b^5 \ub \left. \frac{\partial{\left( - i \Sigma_{22} \right)
}}{\partial p^{0}} \right|_{\ps = m_b}u  =  \label{eq:ward22} \\ \nn
&& -\left[ \ref{fig:wave22}c) + \ref{fig:wave22}d) + 
\ref{fig:wave22}e) \right]_{wave function} 
\eea
Since
\be
\frac{\partial}{\partial p_{0}} \prk = \prk i \gamma^0 \prk
\ee
the derivation of $\Sigma_{22}$ with respect to $p_0$ results in 
the insertion of an extra vertex $i \gamma^{0}$ and an extra 
propagator in every possible way. The resulting diagrams are  
similar to the wave function renormalization parts of 
\ref{fig:wave22}c), \ref{fig:wave22}d) and
\ref{fig:wave22}e) with the $C m_b^5$ vertex replaced by $-C m_b^5 \gamma^0$ 
(see Fig. \ref{fig:ward2}b). Writing out explicitly
\bea
&& C m_b^5 \ub \left. \frac{\partial{\Sigma_{22}}}{\partial p_{0}} 
\right|_{\ps = m_b}u  = 
 -C m_b^5 \cfa (i g)^4 \times \left\{   
\int \ph N_1 \prkd \times \right. \\ \nn
&& \prkd \prkkod \prkod \gk \gko 
 + \int \ph  N_2 \times \\ \nn
&& \prkd \prkkod \prkkod \prkod
\gk \gko  \\ \nn
&& + \int \ph N_3 \prkd \prkkod \prkod \prkod 
\\ \nn 
&&\left.  \gk \gko \right\}
\eea

where $N_1$, $N_2$, $N_3$ are the numerator factors 
\bea
&& N_1 = \ub \gua \prkn \go \prkn \gub \prkkon \gda \prkon \gdb u
\\ \nn
&& N_2 = \ub \gua \prkn \gub \prkkon \go \prkkon \gda \prkon \gdb  u
\\ \nn
&& N_3 = \ub \gua \prkn \gub \prkkon \gda \prkon \go \prkon \gdb u
\eea

Comparing this with diagrams \ref{fig:wave22}c), \ref{fig:wave22}d) and 
\ref{fig:wave22}e) 
\bea 
&& -\ref{fig:wave22}c) = -C m_b^5 \cfa (ig)^4 \int \ph N_{1c} 
\prkd \times \\ \nn 
&&\prkd \prkkod \prkod \gk \gko  \\ \nn
&& -\ref{fig:wave22}d) = -C m_b^5 \cfa (ig)^4 \int \ph N_{2d} \prkd \times \\ \nn
&& \prkkod \prkkod \prkod \gk \gko \\ \nn 
&& -\ref{fig:wave22}e) = -C m_b^5 \cfa (ig)^4 \int \ph N_{3e} \prkd \times \\ \nn
&& \prkkod \prkod \prkod \gk \gko
\eea
with the numerator factors
\bea
&& N_{1c} = \ub \gua \prkn \prkn \gub \prkkon \gda \prkon \gdb u
\\ \nn
&& N_{2d} = \ub \gua \prkn \gub \prkkon \prkkon \gda \prkon \gdb  u
\\ \nn
&& N_{3e} = \ub \gua \prkn \gub \prkkon \gda \prkon \prkon \gdb u
\eea
we see that the proof of equation (\ref{eq:ward22}) is tantamount 
to showing the equality of numerator factors to linear accuracy. 
IR power counting on the integral shows that we have to keep 
the terms in the numerator at most linear in the integration 
variables. To linear accuracy the first numerator pieces, 
$N_1$ and $N_{1c}$ are
\bea 
N_{1} &=& \ub \gua \psm \go \psm \gub \psm 
\gda \psm \gdb u \label{eq:n1} \\ \nn
&-& \ub \gua \ks \go \psm \gub \psm \gda \psm \gdb u \\ \nn
&-& \ub \gua \psm \go \ks \gub \psm \gda \psm \gdb u \\ \nn
&-& \ub \gua \psm \go \psm \gub \left( \ks +\ks_1 \right) \gda \psm
\gdb u \\ \nn
&-& \ub \gua \psm \go \psm \gub \psm \gda \ks_1 \gdb u 
\eea
and 
\bea
N_{1c} &=& \ub \gua \psm \psm 
\gub \psm \gda \psm \gdb u \label{eq:nc}\\ \nn
&-& \ub \gua \ks \psm \gub \psm \gda \psm \gdb u \\ \nn
&-& \ub \gua \psm \ks \gub \psm \gda \psm \gdb u \\ \nn
&-& \ub \gua \psm \psm \gub \left( \ks + \ks_1 \right) \gda \psm \gdb 
u \\ \nn
&-& \ub \gua \psm \psm \gub \psm \gda \ks_1 \gdb u 
\eea
Note that the first terms in equations (\ref{eq:n1}), 
(\ref{eq:nc}) give
the logarithmic infrared sensitive pieces, while the last 4 
terms give the linear infrared sensitive parts. 
Comparing the first terms, it is easy to see
the equality of the logarithmic pieces.

It is also straightforward to see that $\left. N_1 
\right|_{linear}= \left. 
N_{1c} \right|_{linear}$. The only difference in the 
numerators is the presence of the $\go$ factor. 
Since the numerator is at most linear in $k$ and $k_1$, either left
or right to $\go$ there will be only factors of $\psm$ and $\gamma$ 
matrices, which could be easily evaluated using the mass-shell 
condition. Then using that $\go u = u$ (since in rest frame) we 
have effectively replaced the $\go$ by 1, which means the equality of
the two numerator pieces. In the same way one can show that to linear 
accuracy $N_2 = N_{2d}$ and $N_3 =N_{3e}$, thus the
equality of numerators and hence equation (\ref{eq:ward22}) holds 
to linear accuracy. Therefore the logarithmic and linear
infrared sensitive wave function renormalization pieces of the 
$\Sigma_{22}$ group cancel 
\be
\left[ \ref{fig:wave22}a) + \ref{fig:wave22}b) + \ref{fig:wave22}c) 
+ \ref{fig:wave22}d) + \ref{fig:wave22}e) \right]_{
wave function} = 0
\ee

The cancellation of the logarithmic and linear wave function parts of group-4 
can be shown in a similar way. The cancellation of logarithmic and linear
wave function renormalization pieces in group-5 is discussed in the 
Appendix.

The first and third group do not cancel separately, since $\Sigma_{21}$
contains a mass renormalization piece through its inner part, $\Sigma_1$.
This piece will cancel with the corresponding mass counterterm of $\Sigma_{23}$,
and the sum of the two groups again cancel to the accuracy desired.
To show the cancellation of the logarithmic and linear 
infrared sensitive pieces in the wave function renormalization, 
we again use the Ward-like identities in Fig.s 
\ref{fig:ward2}a) and \ref{fig:ward2}c).

At the two-loop level 1-particle reducible diagrams also contribute to 
the wave function renormalization. The relevant diagrams are shown 
in Fig. \ref{fig:1pr}. The contribution of the diagrams (shown in
the Figure) can be simply calculated using the perturbative expansion
of $Z^{1/2}$ (Eq. \ref{wf1}) multiplying each external line in the S-matrix
element. As Fig. \ref{fig:1pr} shows, the wave function renormalization
pieces from the two-loop order 1-particle reducible diagrams cancel.

\section{Cancellation of the Infrared Sensitive Terms Arising from Mass
Renormalization} \label{sec:mass}
Let us now consider the cancellation of the leading infrared sensitive piece (the
one that is linearly divergent in the infrared) arising from the diagrams
involving the mass renormalization of the heavy quark. As for the wave function
case we begin with a discussion of the one loop case to exemplify our method and
then the more complicated two loop example is considered.

\subsection{Cancellation at One Loop} \label{sec:mass1}
As discussed in the introduction, the pole mass of the heavy quark contains a
long distance piece which is proportional linearly to an infrared cutoff,
the first contribution starting at order $\alpha$. In fact, for our subsequent
analysis it is convenient to separate out this piece and write for the pole mass
to one loop order:
\be
m_b~=~\bar{m} + b_1g^2\lambda
\ee
where, as before, $\bar{m}$ is used to denote the short distance or the running mass,
which itself is a power series expansion in $g^2$:
\be
\bar{m}~=~m_0 + a_1g^2.
\ee
 We should emphasize our notation at
this point. Even though one may use a gluon mass as an infrared cutoff at 
the one loop level, this is not a gauge invariant procedure beyond it. 
Thus $\lambda$ refers to some gauge invariant cutoff. Throughout this paper, 
however we will not have any need to specify this infrared cutoff. 
Terms that are linearly divergent in the infrared are identified by power 
counting and cancellation of such terms is shown at the level of the 
corresponding Feynman integrands. Thus for example,
\be
b_1g^2\lambda~\equiv~g^2 C_F \int_{linear}{ d^4k \over (2\pi)^4}{1 \over
k^2+i\epsilon}{1 \over k_0-i\epsilon}.
\ee
The various diagrams contributing to this order are shown in 
Fig. \ref{fig:mass1}.
Fig. \ref{fig:mass1}a) is the bremsstrahlung contribution. 
To find the leading infrared sensitive contribution, we expand the leading 
effective hard vertex to first order in the offshellness, i.e.,
\be
C (\ps-\ks)^5 = C (\ps-\ks-m_b+m_b)^5 =
C m_b^5+ C 5m_b^4(\ps-\ks-m_b) + \cdots
\ee
This is an expansion of the leading effective vertex in  
$(\ps-\ks-m_b) / m_b$ and it is understood that we keep such deviations 
from the mass shell in as much as they
cancel the corresponding small denominators from the propagators. 
Then after this expansion, one of the fermion propagators in the integrand 
of the expression for Fig. \ref{fig:mass1} is cancelled (shown by a slash 
in the Figure) and we are left with an expression which resembles that for 
the self energy. In fact,
\be
\ref{fig:mass1}a)=C 5(\bar{m})^4b_1g^2\lambda \ub u
\ee
Next we have the contributions from the bremsstrahlung from the final state quark
which as discussed earlier is replaced by the effective vertices for the gluon
emission. The two diagrams with single gluon effective vertices of Figs. 
\ref{fig:mass1}b) and \ref{fig:mass1}c) give: 
\be
\ref{fig:mass1}b) + \ref{fig:mass1}c) = 2 \times C(-i5m_b^4) \bar{u} \left(-i
\Sigma_1(p) \right)u
 = -2 \times C 5(\bar{m})^4b_1g^2\lambda \ub u.
\ee
Finally, in the lowest order diagram, Fig. \ref{fig:mass1}d), 
we must express the leading effective vertex in terms of the short distance 
mass,
\be
\ref{fig:mass1}d)~=~Cm_b^5= C(\bar{m} + b_1g^2\lambda)^5 = C \bar{m}^5 \ub u+ 
5 C (\bar{m})^4 b_1g^2\lambda \ub u.
\ee
Adding together all the order $g^2$ terms we see the cancellation of the
linearly infrared divergent pieces to this order when the decay rate is 
expressed in terms of the running mass $\bar{m}$. We next discuss the two 
loop example.

\subsection{Cancellation to Two Loops}

As in the wave function renormalization, we group the two-loop 1 PI 
diagrams according to the five self-energy pieces shown in Fig. 
\ref{fig:selfen}. These groups are now enlarged by new diagrams with the
single and double gluon vertices 
(Fig. \ref{fig:effvertex}). Unlike in the wave function case, where each 
group of graphs cancelled separately, the cancellation of linear  
infrared sensitive pieces in the mass renormalization
involves mixing between the different groups and mass shift terms
from one loop graphs. This is essentially because whereas the wave
function renormalization is multiplicative, the on-shell 
mass renormalization is additive.

By infrared power counting we can isolate the linear 
infrared sensitive
pieces of the self-energy diagrams as
\begin{eqnarray}
\left.\Sigma_1 \right|_{\ps=m_b} &=& a_1 g^2 + b_1 g^2 \lambda 
\label{eq:sigm}\\ \nn
\left.\Sigma_{2i} \right|_{\ps=m_b} &=& a_{2i} g^4 + b_{2i} g^4 \lambda \qquad
i=1 \ldots 5 
\end{eqnarray}
(Note that as before, $\lambda$ denotes symbolically the linearly 
infrared sensitive pieces of
the diagrams identified by power counting, not an actual infrared cutoff 
parameter.)
Define the running mass as
\begin{equation}  
\bar{m} = m_0 + a_1 g^2 + \sum_{i=1}^{5} a_{2i} g^4.
\end{equation}
Then the physical mass in terms of the running mass can be written as
\begin{equation}
m_b = \bar{m} + b_1 g^2 \lambda + \sum_{i=1}^{5} b_{2i} g^4 \lambda.
\end{equation}

\subsubsection{Group-1 and Group-2}

\paragraph{Leading effective vertex and single gluon vertex graphs}

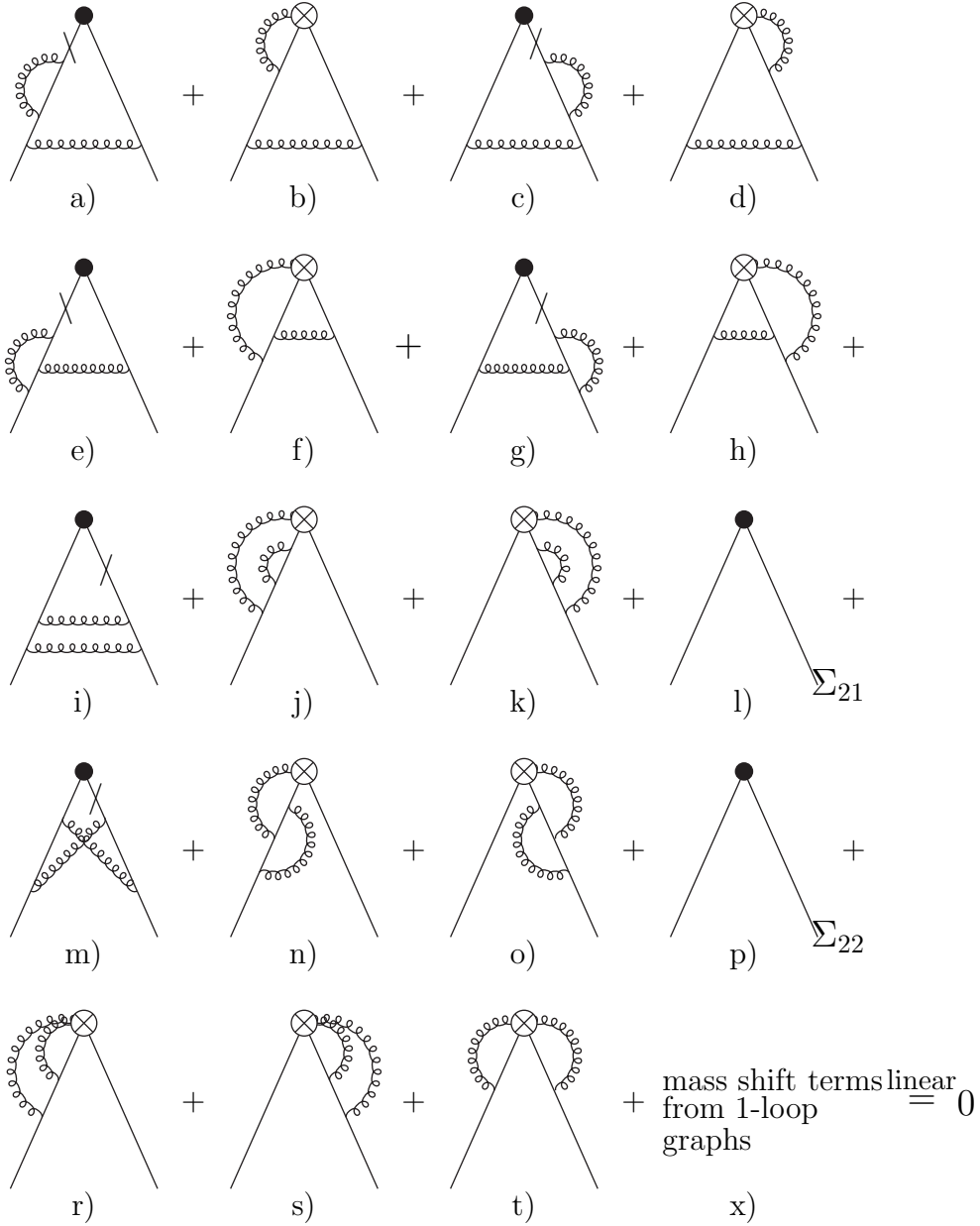
\begin{figure}


\begin{center} \begin{picture}(400,90)(0,0)

\SetScale{7}
\SetWidth{0.071}

\SetOffset(0,12)
\Line(1,1)(5,10)
\Line(9,1)(5,10)
\Vertex(5,10){0.5}
\Gluon(1.88,3)(8.12,3){-0.23}{9}
\GlueArc(3.22,6)(1.7,66,246){-0.23}{9}
\Text(29.7,58.1)[]{$\backslash$}
\Text(35,0)[]{a)}

\Text(75,40)[]{$\large$ $+$}

\SetOffset(84,12)
\Line(1,1)(5,10)
\Line(9,1)(5,10)
\GlueArc(4.31,8.45)(1.7,66,246){-0.23}{9}
\Gluon(1.88,3)(8.12,3){-0.23}{9}
\BCirc(5,10){0.7}
\Line(5.5,9.5)(4.5,10.5) \Line(4.5,9.5)(5.5,10.5)
\Text(35,0)[]{b)}

\Text(75,40)[]{$\large$ $+$}

\SetOffset(168,12)
\Line(1,1)(5,10)
\Line(9,1)(5,10)
\Vertex(5,10){0.5}
\Gluon(1.88,3)(8.12,3){-0.23}{9}
\GlueArc(6.78,6)(1.7,-66,114){-0.23}{9}
\Text(39.67,59.5)[]{$\slash$}
\Text(35,0)[]{c)}

\Text(75,40)[]{$\large$ $+$}

\SetOffset(252,12)
\Line(1,1)(5,10)
\Line(9,1)(5,10)
\Gluon(1.88,3)(8.12,3){-0.23}{9}
\GlueArc(5.7,8.45)(1.7,-66,114){-0.23}{9}
\BCirc(5,10){0.7}
\Line(5.5,9.5)(4.5,10.5) \Line(4.5,9.5)(5.5,10.5)
\Text(35,0)[]{d)}

\end{picture}
\end{center}

\begin{center} \begin{picture}(400,90)(0,0)

\SetScale{7}
\SetWidth{0.071}

\SetOffset(0,12)
\Line(1,1)(5,10)
\Line(9,1)(5,10)
\Vertex(5,10){0.5}
\Gluon(2.55,4.5)(7.45,4.5){-0.23}{9}
\GlueArc(2.655,4.73)(1.7,66,246){-0.23}{9}
\Text(28.14,54.6)[]{$\backslash$}
\Text(35,0)[]{e)}

\Text(75,40)[]{$\large$ $+$}

\SetOffset(84,12)
\Line(1,1)(5,10)
\Line(9,1)(5,10)
\GlueArc(3.86,7.44)(2.8,66,246){-0.23}{13}
\Gluon(3.36,6.32)(6.63,6.32){-0.23}{5}
\BCirc(5,10){0.7}
\Line(5.5,9.5)(4.5,10.5) \Line(4.5,9.5)(5.5,10.5)
\Text(35,0)[]{f)}

\Text(75,40)[]{\large $+$}

\SetOffset(168,12)
\Line(1,1)(5,10)
\Line(9,1)(5,10)
\Vertex(5,10){0.5}
\Gluon(2.55,4.5)(7.45,4.5){-0.23}{9}
\GlueArc(7.345,4.73)(1.7,-66,114){-0.23}{9}
\Text(41.85,54.6)[]{$\slash$}
\Text(35,0)[]{g)}

\Text(75,40)[]{$\large$ $+$}

\SetOffset(252,12)
\Line(1,1)(5,10)
\Line(9,1)(5,10)
\GlueArc(6.14,7.44)(2.8,-66,114){-0.23}{13}
\Gluon(3.36,6.32)(6.63,6.32){-0.23}{5}
\BCirc(5,10){0.7}
\Line(5.5,9.5)(4.5,10.5) \Line(4.5,9.5)(5.5,10.5)
\Text(35,0)[]{h)}

\Text(75,40)[]{$\large$ $+$}
\end{picture}

\end{center}

\begin{center} \begin{picture}(400,90)(0,0)

\SetScale{7}
\SetWidth{0.071}

\SetOffset(0,12)
\Line(1,1)(5,10)
\Line(9,1)(5,10)
\Vertex(5,10){0.5}
\Gluon(1.88,3)(8.12,3){-0.23}{9}
\Gluon(2.55,4.5)(7.45,4.5){-0.23}{7}
\Text(43.57,50.75)[]{$/$}
\Text(35,0)[]{i)}

\Text(75,40)[]{$\large$ $+$}

\SetOffset(84,12)
\Line(1,1)(5,10)
\Line(9,1)(5,10)
\GlueArc(3.86,7.44)(2.8,66,246){-0.23}{13}
\GlueArc(3.86,7.44)(1.1,66,246){-0.23}{5}
\BCirc(5,10){0.7}
\Line(5.5,9.5)(4.5,10.5) \Line(4.5,9.5)(5.5,10.5)
\Text(35,0)[]{j)}

\Text(75,40)[]{$\large$ $+$}

\SetOffset(168,12)
\Line(1,1)(5,10)
\Line(9,1)(5,10)
\GlueArc(6.14,7.44)(2.8,-66,114){-0.23}{13}
\GlueArc(6.14,7.44)(1.1,-66,114){-0.23}{5}
\BCirc(5,10){0.7}
\Line(5.5,9.5)(4.5,10.5) \Line(4.5,9.5)(5.5,10.5)
\Text(35,0)[]{k)}

\Text(75,40)[]{$\large$ $+$}

\SetOffset(252,12)
\Line(1,1)(5,10)
\Line(9,1)(5,10)
\Vertex(5,10){0.5}
\Text(71,7)[]{\large $\Sigma_{21}$}
\Text(35,0)[]{l)}

\Text(75,40)[]{$\large$ $+$}

\end{picture}
\end{center}

\begin{center} \begin{picture}(400,90)(0,0)

\SetScale{7}
\SetWidth{0.071}

\SetOffset(0,12)
\Line(1,1)(5,10)
\Line(9,1)(5,10)
\Vertex(5,10){0.5}
\Gluon(3.88,7.5)(7.88,3.5){-0.23}{9}
\Gluon(2.12,3.5)(6.12,7.5){-0.23}{9}
\Text(39.67,59.5)[]{$\slash$}
\Text(35,0)[]{m)}

\Text(75,40)[]{$\large$ $+$}

\SetOffset(84,12)
\Line(1,1)(5,10)
\Line(9,1)(5,10)
\GlueArc(4.186,8.17)(2,66,246){-0.23}{11}
\GlueArc(3.37,6.34)(2,-114,66){-0.23}{11}
\BCirc(5,10){0.7}
\Line(5.5,9.5)(4.5,10.5) \Line(4.5,9.5)(5.5,10.5)
\Text(35,0)[]{n)}

\Text(75,40)[]{$\large$ $+$}

\SetOffset(168,12)
\Line(1,1)(5,10)
\Line(9,1)(5,10)
\GlueArc(5.894,8.17)(2,-66,114){-0.23}{11}
\GlueArc(6.63,6.34)(2,114,294){-0.23}{11}
\BCirc(5,10){0.7}
\Line(5.5,9.5)(4.5,10.5) \Line(4.5,9.5)(5.5,10.5)
\Text(35,0)[]{o)}

\Text(75,40)[]{$\large$ $+$}

\SetOffset(252,12)
\Line(1,1)(5,10)
\Line(9,1)(5,10)
\Vertex(5,10){0.5}
\Text(71,7)[]{\large $\Sigma_{22}$}
\Text(35,0)[]{p)}

\Text(75,40)[]{$\large$ $+$}

\end{picture}
\end{center}

\begin{center} \begin{picture}(400,90)(0,0)

\SetScale{7}
\SetWidth{0.071}

\SetOffset(0,12)
\Line(1,1)(5,10)
\Line(9,1)(5,10)
\GlueArc(3.86,7.44)(2.8,66,246){-0.23}{13}
\GlueArc(4.3,8.44)(1.7,66,246){-0.23}{9}
\BCirc(5,10){0.7}
\Line(5.5,9.5)(4.5,10.5) \Line(4.5,9.5)(5.5,10.5)
\Text(35,0)[]{r)}

\Text(75,40)[]{$\large$ $+$}

\SetOffset(84,12)
\Line(1,1)(5,10)
\Line(9,1)(5,10)
\GlueArc(6.14,7.44)(2.8,-66,114){-0.23}{13}
\GlueArc(5.69,8.44)(1.7,-66,114){-0.23}{9}
\BCirc(5,10){0.7}
\Line(5.5,9.5)(4.5,10.5) \Line(4.5,9.5)(5.5,10.5)
\Text(35,0)[]{s)}

\Text(75,40)[]{$\large$ $+$}

\SetOffset(168,12)
\Line(1,1)(5,10)
\Line(9,1)(5,10)
\GlueArc(4.186,8.17)(2,66,246){-0.23}{11}
\GlueArc(5.894,8.17)(2,-66,114){-0.23}{11}
\BCirc(5,10){0.7}
\Line(5.5,9.5)(4.5,10.5) \Line(4.5,9.5)(5.5,10.5)
\Text(35,0)[]{t)}

\Text(75,40)[]{$\large$ $+$}

\SetOffset(252,12)
\Text(0,49)[l]{$\large$ mass shift terms }
\Text(0,37)[l]{$\large$ from 1-loop }
\Text(0,25)[l]{$\large$ graphs}
\Text(35,0)[]{x)}

\Text(103,40)[]{\large $=$}
\Text(120,40)[]{\large $0$}
\Text(103,49)[]{\small linear}

\end{picture}
\end{center}

\caption{\setlength{\baselineskip}{0.30in}
Mass renormalization of order $g^4$: group-1 and group-2. 
The linear infrared sensitive mass renormalization pieces of 
the leading effective vertex graphs cancel with the 
corresponding single and double gluon vertex graphs and with $g^4$ 
order mass shift terms from the 1-loop diagrams.}
\label{fig:mass21} 
\end{figure}

The cancellation between the linear infrared sensitive pieces in 
the mass renormalization of the first two group of 
graphs with the corresponding single and double gluon vertex graphs is shown in
Fig. \ref{fig:mass21}. 
As before, the linear infrared sensitive pieces in the mass renormalization
of the diagrams are identified by IR power counting, and will be referred
simply as ``linear mass pieces''  or  ``linear pieces'' of the 
diagrams in the following.
Note that the cancellation involves the order $g^4$ 
mass shift terms from the one-loop diagrams, as will be discussed below. 

Let us first look at the first two graphs of 
Figs. \ref{fig:mass21}, \ref{fig:mass21}a) and \ref{fig:mass21}b). 
The vertex of \ref{fig:mass21}a) is proportional
to $\left(\ps-\ks \right)^5$ (we will always take $k$ the momentum of
the first outgoing gluon). 
As discussed for the 1-loop case, the linear mass pieces of the graphs 
are extracted by
expanding the vertex around mass shell up to linear terms in the integration
variables, identified by power counting.
In the case of \ref{fig:mass21}a) 
\begin{equation}
C \left. \left(\ps -\ks \right)^5 \right|_{mass,~linear} = C 5 m_b^4 \left(
\ps-\ks-m_b \right) + 
 \tif \left(\ps-\ks-m_b \right)^2 \label{eq:levexp} 
\end{equation}
The first term in the vertex expansion, $C m_b^5$, gives 
the wave function renormalization piece of the diagram. The
remaining terms in the vertex expansion contribute to the 
mass renormalization parts of the diagram. These are at least linear 
in the offshellness $(\ps-\ks -m_b)$, 
canceling a propagator of the graph with an extra factor of $i$
left. This cancellation is as was mentioned earlier for the one loop case
denoted by a slash on the corresponding  propagator in the figures.

The vertex of \ref{fig:mass21}b) is a single gluon vertex (Fig. 
\ref{fig:effvertex}), 
its expansion up to terms giving linear infrared sensitivity in the integration 
$$\left. \left[ vertex \right]_{\alpha}^{a} \right|_{linear}=
C g~m_b^4 T^a \left[ 
(\ps-\ks)^4\gamma_{\alpha} + (\ps-\ks)^3 \gamma_{\alpha} (\ps -\ks -\ks_{1})+ 
\right.$$
\begin{equation}
\left. \left. \cdots+ \gamma_{\alpha} (\ps -\ks -\ks_1)^4 \right] 
\right|_{linear}= 
C \frac{5}{i} m_b^4  T^a i g\gamma_{\alpha} + C \frac{10}{i} m_b^3  
T^a (\ps -\ks -m_b)
i g \gamma_{\alpha} 
\end{equation}

Substituting the linearized expressions for the vertices, we find that 
\ref{fig:mass21}a) and \ref{fig:mass21}b) cancel. 
The same type of cancellation applies for the next pair of graphs, 
\ref{fig:mass21}c) and \ref{fig:mass21}d)
\be
\ref{fig:mass21}a)+\ref{fig:mass21}b)=0 \quad \ref{fig:mass21}c) + 
\ref{fig:mass21}d)=0.
\ee

The case of \ref{fig:mass21}e-f) and \ref{fig:mass21}g-h) is a 
little more complicated. The leading effective
vertex is expanded the same way as in (\ref{eq:levexp}), giving 
\bea
\ref{fig:mass21}e) &=& 
\left.  \fif \left(-i \Sigma_{22} \right) \right|_{\ps=m_b} \ub u
-\tif \cfa \ub i g \gua i g \gub \times \label{eq:sgvexpt}\\ \nn 
&& \int \ph \prkko i g \gda \prko i g \gdb \gk \gko u    
\eea  

Here however the expansion of the single gluon vertex to linear IR sensitive
terms in the integration results in 3 terms:
\bea
\left.
\veb \right|_{linear} &=& C g T^b \left[ \left(\ps -\ks \right)^4 \gdb + \left(\ps -\ks \right)^3 
\gdb  \left(\ps - \ks - \ks_1 \right)+ \ldots +\right.\\ \nn
&&\left. \left. \gdb \left(\ps- \ks - \ks_1\right)^4  \right] \right|_{linear}= 
C \frac{5}{i} m_b^4 \gdbt 
+ C \frac{
10}{i} m_b^3 \left(\ps -\ks - m_b \right) \gdbt \\ \nn
&& + C \frac{10}{i} m_b^3 \gdbt \left(\ps -\ks - \ks_1 -m_b \right) 
\eea
From (\ref{eq:sgvexpt}) we get for  \ref{fig:mass21}f)
\bea 
\ref{fig:mass21}f) &=& \left. -\fif \left( -i \Sigma_{22} \right) 
\right|_{\ps=m_b} \ub u + 
\tif \cfa \ub  i g \gua i g \gub \times \\ \nn
&& \int \ph \prkko  i g \gda \prko i g \gdb \gk \gko u + \\ \nn 
&& \tif \cfa \ub i g \gua 
\int \ph \prk \times\\ \nn
&& i g \gub i g \gda \prko i g \gdb \gk \gko u 
\eea 
Unlike the first two pairs of graphs, \ref{fig:mass21}e) and \ref{fig:mass21}f) do not cancel. 
The sum of the two graphs is the last term of \ref{fig:mass21}f).
Calculating for \ref{fig:mass21}g) and \ref{fig:mass21}h) in the same way, we find that 
\bea  
\ref{fig:mass21}g)+\ref{fig:mass21}h) &=& \ref{fig:mass21}e) + 
\ref{fig:mass21}f) \\ \nn
\ref{fig:mass21}e)+\ref{fig:mass21}f)+\ref{fig:mass21}g)+\ref{fig:mass21}h) 
&=& 2 \times \tif \cfa \ub i g \gua \times \\ \nn
&& \int \ph \prk i g \gub i g \gda \prko i g \gdb \gk \gko u 
\eea 

Next consider the diagrams \ref{fig:mass21}i)-l). 
Expanding the vertices around mass shell and keeping linear 
infrared sensitive terms in the integration gives 
\be
\ref{fig:mass21}i) = 
\left. \fif \left(-i \Sigma_{21} \right) \right|_{\ps=m_b} \ub u
\ee
(Note that further expansion of the leading effective vertex would give
a quadratically infrared sensitive piece by IR power counting.)

The single gluon vertex of \ref{fig:mass21}j) is expanded  to linear accuracy as
\bea
\left. \left[ vertex \right]_{\alpha}^{a} \right|_{linear}&=&C g T^a \left[ 
\ps^4\gamma_{\alpha} + \ps^3 \gamma_{\alpha} (\ps-\ks) + 
\left. \cdots+ \gamma_{\alpha} (\ps-\ks)^4 \right] \right|_{linear}= 
\label{eq:sgvexp2p} \\ \nn  
&& C \frac{5}{i} m_b^4  T^a i g~\gamma_{\alpha} + C \frac{
10}{i} m_b^3  T^a i g  \gamma_{\alpha} (\ps -\ks -m_b)  
\eea
Substituting the vertex expansion (\ref{eq:sgvexp2p}), \ref{fig:mass21}j) becomes
\bea
\ref{fig:mass21}j) &=& \left. -\fif \left(-i \Sigma_{21}\right) \right|_
{\ps=m_b} \ub u\\  
&& + \tif \cf \ub i g \gua i g \gub \int
\ph \prkko i g \gdb \prk i g \gda \gk \gko u \nn 
\eea
A similar calculation for \ref{fig:mass21}k) gives
\bea
\ref{fig:mass21}k) &=& \left. -\fif \left(-i \Sigma_{21}\right) 
\right|_{\ps=m_b} \ub u \\
&& + \tif \cf \ub i g \gua \int \ph \prk i g \gub \prkko i g \gdb i g \gda  \gk
\gko u \nn
\eea
Diagram \ref{fig:mass21}l) is the mass shift term of the tree level 
amplitude corresponding to $\Sigma_{21}$
\be
\ref{fig:mass21}l)= C \left. \ub \ps^5 u \right|_{\Sigma_{21}}= \left. 
C m_b^5 \right|_{\Sigma_{21}} \ub u = C 5 \bar{m}^4 g^4 b_{21} \lambda \ub u  
\ee
Adding the 4 graphs and using that $\Sigma_{21}=a_{21}+ b_{21} g^4 \lambda$ 
we arrive at
\bea
&& \ref{fig:mass21}i)+\ref{fig:mass21}j)+\ref{fig:mass21}k)+
\ref{fig:mass21}l) =\\ \nn
&& \tif \cf \ub i g \gua i g \gub \int
\ph \prkko i g \gdb \prk i g \gda \gk \gko u + \\ \nn
&& \tif \cf \ub i g \gua \int \ph \prk i g \gub \prkko i g \gdb i g \gda  \gk
\gko u 
\eea

The calculation for the next 4 graphs, \ref{fig:mass21}m-p) is similar to 
\ref{fig:mass21}i-l. Note
however that in the leading effective vertex expansion of \ref{fig:mass21}m) we must keep
the first 2 terms, unlike in the case of \ref{fig:mass21}i), resulting in an extra term.
Apart from this, the difference is only in the gamma matrix and color 
structure. For the sum we find 
\bea
&& \ref{fig:mass21}m)+\ref{fig:mass21}n)+\ref{fig:mass21}o)+
\ref{fig:mass21}p) =\\ \nn 
&& -\tif \cfa \ub i g \gua \times \\ \nn 
&& \int \ph \prk i g \gdb i g \gda \prko i g \gub \gk \gko u \\ \nn 
&& +  \tif \cfa \ub ig \gua \times \\ \nn 
&& \int \ph \prk i g \gub \prkko  i g \gda i g \gdb \gk \gko u \\ \nn 
&& + \tif \cfa \ub i g \gua i g \gub \times \\ \nn 
&& \int \ph \prkko \gdai  \prko i g \gdb \gk \gko u
\eea 

Summarizing the leading and single gluon effective vertex diagrams
contribute with
\bea
&& \ref{fig:mass21}a)+\cdots+\ref{fig:mass21}p)= \label{eq:esvsum} \\ \nn
&& \tif \left\{ \cf \ub i g \gua i g \gub \int
\ph \prkko i g \gdb \prk i g \gda \gk \gko u \right.  \\ \nn
&+& \cfa \ub i g \gua i g \gub \int \ph \prkko i g \gda \prko 
i g \gdb \gk \gko u \\ \nn
&+& \cf \ub i g \gua \int \ph \prk i g \gub \prkko i g \gdb i g \gda  \gk
\gko u \\ \nn
&+& \cfa \ub ig \gua \int \ph \prk i g \gub \prkko i g \gda i g \gdb \gk \gko u 
 \\ \nn
&+&\left. \cfa \ub i g \gua \int \ph \prk i g \gdb i g 
\gda \prko i g \gub \gk \gko u \right\}   
\eea

\paragraph{Double vertex graphs}

Since we keep infrared sensitive terms to linear accuracy, power 
counting shows that the double gluon vertex need not be
expanded any further, and so it is simplified to 
\be
\left[double~vertex \right]_{\alpha, \beta}^{a,b} = 
C 10 g^2 m_b^3 \left( \gda \gdb T^a T^b + \gdb \gda T^b T^a \right)
\ee
From the above,
\bea 
\ref{fig:mass21}r) &=& 
- \tif \cf \ub \gubi \guai \times \\ \nn
&& \int \ph \prkko \gdai \prko \gdbi \gk \gko u \\ \nn
&& - \tif \cfa \ub \guai \gubi \times \\ \nn
&& \int \ph \prkko \gdai \prko \gdbi \gk \gko u
\eea
\bea 
\ref{fig:mass21}s) &=& 
- \tif \cf \ub \guai \times \\ \nn 
&& \int \ph \prk \gubi \prkko \gdbi \gdai \gk \gko u\\ \nn
&& - \tif \cfa \ub \guai \times \\ \nn
&& \int \ph \prk \gubi \prkko \gdai \gdbi \gk \gko u
\eea
\bea
\ref{fig:mass21}t) &=& 
-\tif \cf \ub \guai \times \\ \nn
&& \int \ph \prk \gdai \gdbi \prko \gubi \gk \gko u \\ \nn
&& -\tif \cfa \ub \guai \times \\ \nn
&& \int \ph \prk \gdbi \gdai \prko \gubi \gk \gko u 
\eea
The sum of the linear infrared sensitive pieces of the 
double vertex graphs is therefore
\bea
&&\ref{fig:mass21}r)+\ref{fig:mass21}s)+\ref{fig:mass21}t)= \label{eq:dvsum}
 \\ \nn
&-& \tif \left\{ 
\cf \ub \gubi \guai  
\int \ph \prkko \gdai \prko \gdbi \gk \gko u \right. \\ \nn
&+& \cfa \ub \guai \gubi 
\int \ph \prkko \gdai \prko \gdbi \gk \gko u \\ \nn
&+& \cf \ub \guai    
\int \ph \prk \gubi \prkko \gdbi \gdai \gk \gko u\\ \nn
&+& \cfa \ub \guai 
\int \ph \prk \gubi \prkko \gdai \gdbi \gk \gko u \\ \nn
&+&\cf \ub \guai \int \ph \prk \gdai \gdbi \prko \gubi \gk \gko u \\ \nn
&+& \left. \cfa \ub \guai \int \ph \prk \gdbi \gdai \prko \gubi \gk \gko u
\right\}   
\eea
Adding (\ref{eq:esvsum}) and (\ref{eq:dvsum}) the sum of the leading, single,
and double gluon effective vertex diagrams gives
\bea
&& \ref{fig:mass21}a)+\cdots+\ref{fig:mass21}t) = \\ \nn 
&& -\tif \cf \ub \guai \int \ph \prk \gdai \gdbi \prko \gubi \gk \gko u = \\ \nn
&& -\tif \left. \left( -i \Sigma_1 \right|_{\ps=m_b} \right)^2 \ub u
\eea
Using power counting, the linear term of the sum is 
(with the use of equation (\ref{eq:sigm}))
\be 
\ref{fig:mass21}a)+\cdots+\ref{fig:mass21}t) = 
C 20 \mb^3  a_1 b_1 g^4 \lambda \ub u
\label{eq:v21sum} 
\ee

\paragraph{Mass-shift terms from 1-loop graphs}

The 1-loop diagrams (Fig. \ref{fig:mass1}) also contribute to the order $g^4$ 
cancellations via their mass shift terms. Using the results of section 
\ref{sec:mass1}, the
linear mass renormalization pieces of the 1-loop graphs to order $g^4$ are
\bea
\ref{fig:mass1}a)+\ref{fig:mass1}b)+\ref{fig:mass1}c) 
&=& \left. -C 5 m_b^4 \left. \Sigma \right|_{\ps =m_b} \ub u \right|_{linear}\\ \nn
&=&-C \left. 5  \left( \mb + a_1 g^2 + b_1 g^2 \lambda \right)^4
\left( a_1 g^2 +b_1 g^2 \lambda \right) \ub u \right|_{linear} \\ \nn
&=& - C 5  \mb^4 b_1 g^2 \lambda \ub u - C 40 \mb^3 a_1 b_1 g^4 \lambda \ub u 
\eea
\bea
\ref{fig:mass1}d) &=& \left. C m_b^5  \ub u \right|_{linear}=  
C \left. \left(\mb +a_1 g^2 + b_1 g^2 \lambda + \sum_{i=1}^4 a_{2i} g^4+
 \sum_{i=1}^4 b_{2i} g^4 \lambda \right)^5 \ub u \right|_{linear} \\ \nn
&=&C  5 \mb^4 b_1 g^2 \lambda \ub u + 
C 5 \mb^4 \sum_{i=1}^4 b_{2i} g^4 \lambda  \ub u +
C 20 \mb^3 a_1 b_1 g^4 \lambda \ub u 
\eea 
As we have shown before, the order $g^2$  terms of the 1-loop diagrams cancel.
The mass shift terms due to the order $g^4$  self-energy corrections were
already included in the calculation. 
The order $g^4$ terms we have not included so far are the ``cross terms'', 
those proportional to $a_1 b_1$. These are the terms we denoted with the 
expression ``mass shift terms'' on Fig. \ref{fig:mass21}x), summing up to
\be 
\ref{fig:mass21}x) = -C 20 \mb^3  a_1 b_1 g^4 \lambda \ub u
\ee
This contribution cancels the remaining linear term of the 2-loop vertex 
diagrams of group-1 and group-2,
(\ref{eq:v21sum}), so the linear infrared sensitive terms in 
the mass renormalization of the first two group of 
diagrams with the mass shift terms from the 1-loop diagrams cancel
\be
\ref{fig:mass21}a)+\cdots+\ref{fig:mass21}x) = 0.
\ee

\subsubsection{Group-3 and the group of 1-particle reducible diagrams}

\begin{figure}
\begin{center} \begin{picture}(400,90)(0,0)

\SetScale{7}
\SetWidth{0.071}
\SetOffset(0,12)

\Line(1,1)(5,10)
\Line(9,1)(5,10)
\Vertex(5,10){0.5}
\Gluon(1.88,3)(8.12,3){-0.23}{9}
\Text(23.8469,44.8966)[]{\Large $\star$}
\Text(46.1531,44.8966)[]{$\slash$}
\Text(75,40)[]{$\large$ $+$}
\Text(35,0)[]{a)}

\SetOffset(84,12)
\Line(1,1)(5,10)
\Line(9,1)(5,10)
\GlueArc(3.86,7.44)(2.8,66,246){-0.23}{13}
\Text(27.028,52.09)[]{\Large $\star$}
\BCirc(5,10){0.7}
\Line(5.5,9.5)(4.5,10.5) \Line(4.5,9.5)(5.5,10.5)
\Text(35,0)[]{b)}

\Text(75,40)[]{$\large$ $+$}

\SetOffset(168,12)
\Line(1,1)(5,10)
\Line(9,1)(5,10)
\Vertex(5,10){0.5}
\Gluon(1.88,3)(8.12,3){-0.23}{9}
\Text(46.1531,44.8966)[]{\Large $\star$}
\Text(23.8469,44.8966)[]{$\backslash$}
\Text(35,0)[]{c)}

\Text(75,40)[]{$\large$ $+$}

\SetOffset(252,12)
\Line(1,1)(5,10)
\Line(9,1)(5,10)
\GlueArc(6.14,7.44)(2.8,-66,114){-0.23}{13}
\BCirc(5,10){0.7}
\Line(5.5,9.5)(4.5,10.5) \Line(4.5,9.5)(5.5,10.5)
\Text(42.972,52.09)[]{\Large $\star$}
\Line(5.5,9.5)(4.5,10.5) \Line(4.5,9.5)(5.5,10.5)
\Text(35,0)[]{d)}

\Text(75,40)[]{$\large$ $+$}
\end{picture}
\end{center}

\begin{center} \begin{picture}(400,90)(0,0)

\SetScale{7}
\SetWidth{0.071}

\SetOffset(0,12)
\Line(1,1)(5,10)
\Line(9,1)(5,10)
\GlueArc(1.8122,2.827)(1.5,66,246){-0.23}{7}
\Gluon(3,5.55)(7,5.55){-0.23}{5}
\Vertex(5,10){0.5}
\Text(28.14,54.6)[]{$\backslash$}
\Text(35,0)[]{e)}

\Text(75,40)[]{$\large$ $+$}

\SetOffset(84,12)
\Line(1,1)(5,10)
\Line(9,1)(5,10)
\Vertex(5,10){0.5}
\Gluon(3,5.55)(7,5.55){-0.23}{5}
\GlueArc(8.2,2.827)(1.5,-66,114){-0.23}{7}
\Text(28.14,54.6)[]{$\backslash$}
\Text(35,0)[]{f)}

\Text(90,40)[]{\large $+~\cdots~+$}

\SetOffset(188,12)
\Line(1,1)(5,10)
\Line(9,1)(5,10)
\Vertex(5,10){0.5}
\Text(71,7)[]{\large $\Sigma_{23}$}
\Text(35,0)[]{g)}

\Text(75,40)[]{$\large$ $+$}

\end{picture}

\end{center}

\begin{center} \begin{picture}(400,90)(0,0)

\SetScale{7}
\SetWidth{0.071}

\SetOffset(0,12)
\Line(1,1)(5,10)
\Line(9,1)(5,10)
\GlueArc(4.30,8.45)(1.7,66,246){-0.23}{7}
\GlueArc(1.8122,2.827)(1.5,66,246){-0.23}{7}
\BCirc(5,10){0.7}
\Line(5.5,9.5)(4.5,10.5)\Line(4.5,9.5)(5.5,10.5)
\Text(35,0)[]{h)}

\Text(75,40)[]{$\large$ $+$}

\SetOffset(84,12)
\Line(1,1)(5,10)
\Line(9,1)(5,10)
\GlueArc(5.69,8.45)(1.7,-66,114){-0.23}{7}
\GlueArc(8.2,2.827)(1.5,-66,114){-0.23}{7}
\BCirc(5,10){0.7}
\Line(5.5,9.5)(4.5,10.5) \Line(4.5,9.5)(5.5,10.5)
\Text(35,0)[]{i)}

\Text(85,40)[]{\large $+~\cdots~+$}

\SetOffset(188,12)
\Line(1,1)(5,10)
\Line(9,1)(5,10)
\GlueArc(4.30,8.45)(1.7,66,246){-0.23}{7}
\GlueArc(7,5.5)(1.7,-66,114){-0.23}{7}
\BCirc(5,10){0.7}
\Line(5.5,9.5)(4.5,10.5) \Line(4.5,9.5)(5.5,10.5)
\Text(35,0)[]{j)}

\Text(75,40)[]{$\large$ $+$}

\SetOffset(272,12)
\Line(1,1)(5,10)
\Line(9,1)(5,10)
\GlueArc(3,5.5)(1.7,66,246){-0.23}{7}
\GlueArc(5.69,8.45)(1.7,-66,114){-0.23}{7}
\BCirc(5,10){0.7}
\Line(5.5,9.5)(4.5,10.5) \Line(4.5,9.5)(5.5,10.5)
\Text(35,0)[]{k)}

\Text(75,40)[]{\large $+~\cdots~+$}
\Text(108,40)[]{\large $=$}
\Text(125,40)[]{\large $0$}
\Text(108,49)[]{\small linear}
\end{picture}

\end{center}
\caption{\setlength{\baselineskip}{0.30in}
Mass renormalization of order $g^4$: cancellation of 
linear mass terms between group-3 and the 1-particle 
reducible diagrams.
The ellipsis denote inclusion of mass counterterms.}
\label{fig:mass23}
\end{figure}
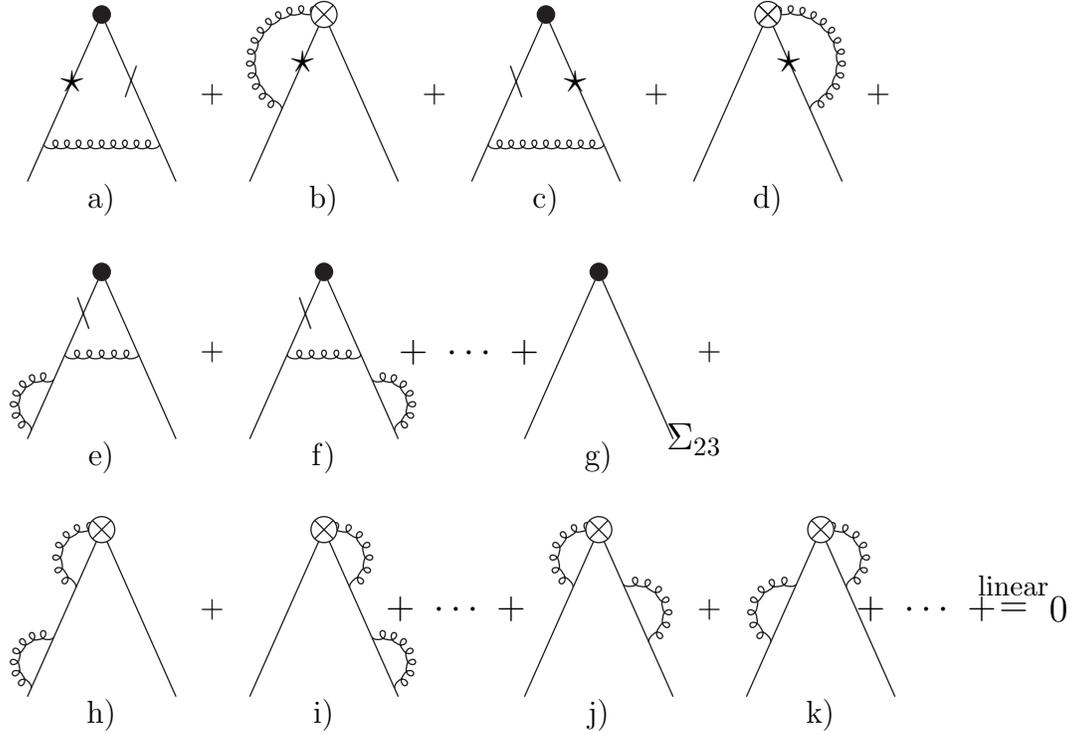

The linear mass renormalization parts of the group-3 diagrams cancel with the
corresponding pieces of 1-particle reducible graphs, as shown in Fig. 
\ref{fig:mass23}. 

Expanding the leading effective vertex of \ref{fig:mass23}a) as in 
(\ref{eq:levexp}), and the single gluon vertex of 
\ref{fig:mass23}b) as in (\ref{eq:sgvexp2p}) to linear accuracy 
it is easy to see that 
\ref{fig:mass23}a) and 
\ref{fig:mass23}b) cancel. The next two pair of graphs 
\ref{fig:mass23}c) and 
\ref{fig:mass23}d) cancel
similarly: 
\be
\ref{fig:mass23}a) + 
\ref{fig:mass23}b) = 0 \quad 
\ref{fig:mass23}c) + 
\ref{fig:mass23}d) = 0
\ee
The mass shift term of the tree amplitude, corresponding to $\Sigma_{23}$,
\ref{fig:mass23}g)
\be 
\ref{fig:mass23}g)=\left. C m_b^5 \right|_{\Sigma_{23}} \ub u= 
5 \mb^4 b_{23} g^4 \lambda  \ub u
\label{eq:sum23} \ee
cancels with the remainder of the one-particle reducible group, as we will
show below. 
For the 1-particle reducible diagrams the leading effective vertex of 
\ref{fig:mass23}e), 
\ref{fig:mass23}f) and the corresponding mass counterterm graphs is expanded to 
first order in the offshellness
\be 
\left. \left( \ps -\ks \right)^5  \right|_{mass,~linear} = 
C 5 m_b^4 \left( \ps -\ks -m_b \right) \label{eq:vilexp}
\ee
Keeping the second term as in 
(\ref{eq:levexp}) would result in a quadratic infrared sensitive term by 
IR power counting. From the vertex expansion
(\ref{eq:vilexp}) we get for the linear mass pieces of the diagrams 
\be 
\ref{fig:mass23}e)+
\ref{fig:mass23}f)+\cdots= - \fif \left(Z^{-1}-1 \right)_{g^2} \left. 
\left(-i \Sigma_{1}\right) \right|_{\ps=m_b} \ub u
\ee
The ellipsis in the formula denote the inclusion of the mass counterterm 
diagrams, as in Fig. \ref{fig:mass23}. Here we used the one-loop Ward-like
identity (Fig. \ref{fig:ward1}) and the perturbative expansion of 
$Z^{1/2}$, Eq. (\ref{wf1}).

For the remaining one-particle reducible diagrams 
\ref{fig:mass23}h-k) the expansion
of the single gluon vertex up to linear infrared sensitive terms in the 
integration gives one term
\be 
\left. \left[ vertex \right]_{\alpha}^{a} \right|_{linear} = 
C \frac{5}{i} m_b^4 \gdat  \label{eq:siiexp}
\ee
Substituting the vertex expansion (\ref{eq:siiexp}), we find that
\be
\ref{fig:mass23}h)+
\ref{fig:mass23}i)+\cdots = 
\ref{fig:mass23}j)+
\ref{fig:mass23}k)+\cdots = 
\fif \left(Z^{-1}-1 \right)_{g^2} \left. \left(-i \Sigma_{1}\right) \right|_{
\ps = m_b} \ub u. 
\ee
Thus the sum of the linear pieces of the 1-particle reducible graphs is
\bea
&& \ref{fig:mass23}e)+\cdots \ref{fig:mass23}k)+\cdots = 
\left. \fif \left(Z^{-1}-1 \right)_{g^2} \left(-i \Sigma_{1}\right) 
\right|_{\ps=m_b, linear}  \ub u \label{eq:irsum}\\ \nn 
&&= C 5 \mb^4 \left(Z^{-1}-1 \right)_{g^2} b_1 g^2 \lambda  \ub u
\eea
The sum (\ref{eq:irsum}) cancels the leftover $\Sigma_{23}$
mass shift piece, (\ref{eq:sum23}), since by the one loop Ward-like identity
(Fig. \ref{fig:ward1})
\be
b_{23} g^2 = -b_1 \left(Z^{-1}-1 \right)_{g^2} 
\ee 
and Eq. (\ref{eq:sum23}) (Fig. \ref{fig:mass23}g)) becomes
\be 
\ref{fig:mass23}g) = C 5 \mb^4 b_{23} g^4 \lambda \ub u= 
- C 5 \mb^4 b_1 g^2 \left(Z^{-1}-1 \right)_{g^2} 
\lambda \ub u,
\ee 
Hence all linear mass renormalization pieces of the group-3 and the 1-particle
reducible diagrams cancel:
\be
\ref{fig:mass23}g)+
\ref{fig:mass23}e)+\cdots 
\ref{fig:mass23}k)+\cdots =0 
\ee

\subsubsection{Group-4}

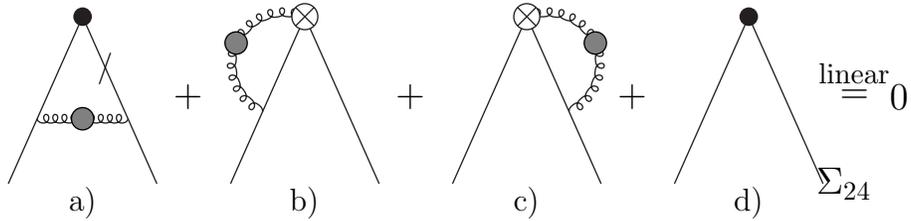
\begin{figure}
\begin{center} \begin{picture}(360,90)(0,0)

\SetScale{7}
\SetWidth{0.071}

\SetOffset(0,12)
\Line(1,1)(5,10)
\Line(9,1)(5,10)
\Vertex(5,10){0.5}
\Gluon(2.55,4.5)(7.45,4.5){-0.23}{9}
\GCirc(5,4.5){0.6}{0.5}
\Text(43.57,50.75)[]{$/$}
\Text(35,0)[]{a)}

\Text(75,40)[]{\large $+$}

\SetOffset(84,12)
\Line(1,1)(5,10)
\Line(9,1)(5,10)
\GlueArc(3.86,7.44)(2.8,66,246){-0.23}{13}
\GCirc(1.30,8.58){0.6}{0.5}
\BCirc(5,10){0.7}
\Line(5.5,9.5)(4.5,10.5) \Line(4.5,9.5)(5.5,10.5)
\Text(35,0)[]{b)}

\Text(75,40)[]{\large $+$}

\SetOffset(168,12)
\Line(1,1)(5,10)
\Line(9,1)(5,10)
\GlueArc(6.14,7.44)(2.8,-66,114){-0.23}{13}
\GCirc(8.69,8.58){0.6}{0.5}
\BCirc(5,10){0.7}
\Line(5.5,9.5)(4.5,10.5) \Line(4.5,9.5)(5.5,10.5)
\Text(35,0)[]{c)}

\Text(75,40)[]{\large $+$}

\SetOffset(252,12)
\Line(1,1)(5,10)
\Line(9,1)(5,10)
\Vertex(5,10){0.5}
\Text(71,7)[]{\large $\Sigma_{24}$}
\Text(35,0)[]{d)}

\Text(75,40)[]{\large $=$}
\Text(92,40)[]{\large $0$}
\Text(75,49)[]{\small linear}
\end{picture}

\end{center}
\caption{\setlength{\baselineskip}{0.30in}
Mass renormalization of order $g^4$: cancellation of linear 
pieces in group-4} 
\label{fig:mass24}

\end{figure}

The cancellation within group-4 (Fig. \ref{fig:mass24}) proceeds in the 
same way as for the 1-loop diagrams (see Fig. \ref{fig:mass1}, 
and section \ref{sec:mass1}). 

\subsubsection{Group-5}

Fig. \ref{fig:mass25} shows the cancellation between the group-5 diagrams.
We can calculate the linear parts of the first 4 diagrams \ref{fig:mass25}a-d) expanding 
the leading effective vertex to second order and the single gluon
vertex to first order in the offshellness. A straightforward calculation gives 
\be 
\ref{fig:mass25}a) + \ref{fig:mass25}b) = 0 \quad \ref{fig:mass25}c) +\ref{fig:mass25}d) = 0
\ee

The case of the next 4 diagrams, \ref{fig:mass25}e-h) is more interesting. \ref{fig:mass25}e), the 
mass-shift piece of the tree amplitude corresponding to $\Sigma_{25}$ is 
\be
\ref{fig:mass25}e) = \left. C m_b^5 \right|_{\Sigma_{25}} \ub u  = 
5 C \mb^4 b_{25} g^4 \lambda \ub u
\ee
Denoting the triple gluon vertex with 
$S^{abc}_{\alpha \beta \delta}$, 
the next diagram, \ref{fig:mass25}f) gives
\bea
\ref{fig:mass25}f) &=& \ub \guat \int \phbc \prbc \times \\ \nn
&& [vertex]^{\beta,b} \prc \guct S^{abc}_{\alpha \beta \delta} \gkb \gkc \gkbc 
\\ \nn
\sv &=& g f^{abc} \left[g_{\alpha \beta} \left( k_a - k_b \right)_{\delta}+
g_{\beta \delta} \left( k_b - k_c \right)_{\alpha}+
g_{\delta \alpha} \left( k_c - k_a \right)_{\beta} \right]
\eea

The single gluon vertex is expanded to linear accuracy as follows:
\bea
\left. \left[vertex \right]^{\beta,b} \right|_{linear} &=& C \frac{5}{i} m_b^4 
\gubt + C \frac{
10}{i} m_b^3 \left( \ps -\ks_b -\ks_c -m_b \right) \gubt \\ \nn
&& + C \frac{
10}{i} m_b^3 \gubt \left( \ps -\ks_c -m_b \right) 
\eea

Substituting the linearized vertex, \ref{fig:mass25}f) gives 3 terms
\bea
\ref{fig:mass25}f) &=& -\fif \left. \left(-i \Sigma_{25} \right) 
\right|_{\ps=m_b} \ub u
+ \tif \ub \guat \gubt \times \\ \nn
&& \int \phbc \prc \guct \sv \gkb \gkc \gkbc u + \\ \nn
&& \tif \ub \guat \int \phbc \prbc \gubt \guct  \sv \gkb \gkc \gkbc u
\eea

To get the linear terms of \ref{fig:mass25}g) and \ref{fig:mass25}h), we need to keep only the first 
term in the expansion of the double gluon vertex. Thus
\bea
\ref{fig:mass25}g) &=& \tif g^2 \ub \left[ \gua \gub T^a T^b + \gub \gua T^b T^a \right] 
\times \\
\nn
&& \int \phbc \prc \guct \sv \gkb \gkc \gkbc u
\eea
and 
\bea
\ref{fig:mass25}h) &=& \tif g^2 \ub \guat \times \\ \nn
&& \int \phbc \prbc \left[\gub \guc T^b T^c +
 \guc \gub T^c T^b \right] \sv \gkb \gkc \gkbc u  
\eea
Using that $\Sigma_{25}=a_{25} g^4 + b_{25} g^4 \lambda$, the sum of the
group-5 graphs gives
\bea
&& \ref{fig:mass25}a)+\cdots+ \ref{fig:mass25}h) = \\ \nn
&& - \tif \ub \guat \int \phbc \prbc \guct \gubt \sv \gkb \gkc \gkbc u \\ \nn
&& - \tif \ub \gubt \guat \int \phbc \prc \guct \sv \gkb \gkc \gkbc u
\eea 
This remainder can be shown to be zero (up to linear accuracy) 
by renaming variables for example in the first integral for
$k'_c=(k_b +k_c)$ and $k'_b=-k_c$. Therefore the linear mass 
renormalization pieces of group-5 diagrams cancel.

\section{Summary and Conclusions.} \label{sec:summary}
In this paper we have shown by explicit perturbative calculations upto the second
loop order, that when the inclusive decay width of a heavy quark is expressed in
terms of the short distance mass, there is complete accord with the operator product
expansion and the leading power corrections are of order $1/m_b^2$. This is the
first such calculation to the second loop order that checks the dependence on
the infrared cutoff to linear accuracy for the non-abelian theory. 
Because of the
non-abelian interactions there could have been potential problems due to the
singularity in the $t$ channel for processes involving the forward scattering of
soft gluons. However we see explicitly that these do not hinder the cancellation
of the infrared sensitive terms to linear accuracy.
 
As emphasized earlier, the above result
is true not just for the semileptonic decay considered in this paper but for any
inclusive decay, like for example the radiative one. Such a cancellation  was
linked to the KLN theorem which had been suggested earlier
\cite{az1} to be the general principle behind  this for inclusive enough
observables.  Since for the inclusive decay, there is only a single particle in
the initial state, our calculations enjoyed several simplifications. It is known
\cite{bloch} that for at least two colored particles in the initial state there is
a breakdown of the Bloch Nordsieck mechanism in perturbation theory. Thus it would
be interesting to  investigate the fate of the $1/Q$ corrections for the Drell Yan
process.  For this process the initial state is no longer trivial and in
addition furthur new complications arise in higher loops due to the mixing of soft
and collinear singularities \cite{azm}. The $1/Q$ corrections are known \cite{bb}
to cancel to  one loop order and the situation at higher loops is an open
question.The techniques developed in  this paper would be useful for this study as
well, and we hope to return to this question in a future publication.

\section{Acknowledgements} This work was supported in part by the US Department
of Energy.

\appendix
\section{Cancellation of Infrared Sensitive Terms to 
Linear Accuracy in the Wave Function Renormalization:
group-5}

For the last group of diagrams (Fig. \ref{fig:wave25}) the
presence of the non-abelian vertex makes the algebra 
a little more complicated. 
The calculation proceeds essentially along the same lines as 
for the second group of diagrams (see section \ref{sec:wave22}).
We would like to show that for the logarithmically and linearly infrared 
sensitive parts (referred to as ``logarithmic and linear pieces''
in the following) of the diagrams of Fig. \ref{fig:wave25}
\bea
&& \left[ \ref{fig:wave25}a) + \ref{fig:wave25}b) 
\right]_{wave function}=
-C m_b^5 \zet_{g^4, \Sigma_{25}} \ub u \\ \nn &&
= -\left[ \ref{fig:wave25}c) + \ref{fig:wave25}d) 
\right]_{wave function}
\eea
As before, the logarithmically and linearly infrared sensitive
pieces of the integrals are isolated by IR power counting.  
In the rest frame of the heavy quark $p=(m_b,0,0,0)$, and
\be
\zet_{g^4, \Sigma_{25}} = -\frac{\ub \left.
\frac{\partial{\Sigma_{25}}}{\partial p_{\nu}} \right|_{\ps = m_b}
u }{\ub \gamma^{\nu} u} =-\frac{\ub \left.
\frac{\partial{\Sigma_{25}}}{\partial p_{0}} \right|_{\ps = m_b}
u }{\ub u}
\ee
so we would like to show that for the logarithmic and linear
terms in the integration
\bea
&& C m_b^5 \ub \left. \frac{\partial{\Sigma_{25}}}{\partial p_{0}}
\right|_{\ps = m_b}u  =
i C m_b^5 \ub \left. \frac{\partial{\left( - i \Sigma_{25} \right)
}}{\partial p_{0}} \right|_{\ps = m_b}u  \label{eq:ward25} = 
-\left[ \ref{fig:wave25}c) + \ref{fig:wave25}d) \right]_{wave function}
\eea
As before, the derivation of $\Sigma_{25}$ with respect to $p_0$
results in an insertion of an extra vertex $i \gamma^{0}$ 
and an extra propagator in all possible way. The resulting
diagrams are similar to \ref{fig:wave25}c) and 
\ref{fig:wave25}d) with the $C m_b^5$ vertex 
replaced by $-C m_b^5 \gamma^0$ (shown in Fig. \ref{fig:ward2}).
Writing out explicitly
\bea 
&&  C m_b^5 \ub \left. \frac{\partial{\Sigma_{25}}}{\partial p_{0}}
\right|_{\ps = m_b}u  = -C m_b^5 \ca (i g)^4 \times \left\{ \int 
\phbc N_1 \times \right. \\ \nn
&& \left[\prdbc \right]^2 \prdc \gkc \gkb \gkbc + 
\int \phbc N_2 \times \\ \nn
&&\left.  \prdbc \left[ \prdc \right]^2 \gkc \gkb \gkbc \right\}
\eea
Denoting the triple gluon vertex by $S_{\alpha \beta \delta}$
(where the color factor and the coupling are taken out) and 
$k_a = -( k_b + k_c )$
\bea
N_1 &=& \ub \gua \left(\ps + \ks_a + m \right) \go \left(
\ps + \ks_a + m \right) \gub \left( \ps -\ks_c +m \right) \guc
S_{\alpha \beta \delta} u \\
N_2 &=& \ub \gua \left(\ps + \ks_a + m \right) \gub 
\left(\ps - \ks_c + m \right) \go \left(\ps - \ks_c + m \right)
\guc S_{\alpha \beta \delta} u \nn \\
S_{\alpha \beta \delta} &=& g_{\alpha \beta} 
\left( k_a - k_b \right)_{\delta} + g_{\beta \delta} 
\left( k_b -k_c \right)_{\alpha} + g_{\delta \alpha} \left(
k_c - k_a \right)_{\beta} \nn
\eea
Comparing this to diagrams \ref{fig:wave25}c) and \ref{fig:wave25}d)
\bea
&&-\ref{fig:wave25}c) = -C m_b^5 \ca \left(i g \right)^4 \int \phbc N_{1c}
\times \\
&& \left[\prdbc \right]^2 \prdc \gkc \gkb \gkbc \nn \\
&&-\ref{fig:wave25}d) = -C m_b^5 \ca \left(i g \right)^4 \int \phbc N_{2d}
\times \nn \\
&& \prdbc \left[ \prdc \right]^2 \gkc \gkb \gkbc \nn
\eea
with the numerator factors
\bea
N_{1c} &=& \ub \gua \left(\ps + \ks_a + m \right) \left(
\ps + \ks_a + m \right) \gub \left( \ps -\ks_c +m \right) \guc
S_{\alpha \beta \delta} u \\
N_{2d} &=& \ub \gua \left(\ps + \ks_a + m \right) \gub 
\left(\ps - \ks_c + m \right) \left(\ps - \ks_c + m \right)
\guc S_{\alpha \beta \delta} u \nn
\eea
we have to show $N_1=N_{1c}$ and $N_2=N_{2d}$ to linear order 
in the integration.
By power counting we must keep terms to the order $k$ in the numerators
in addition to the triple gluon vertex $k$ dependence. 
The logarithmic terms arise
dropping the k dependence everywhere in the numerators except in the 
triple gluon vertex $S_{\alpha \beta \delta}$. A simple calculation shows that
the numerator factors for the logarithmic terms are zero.  
For the linear terms, a little more calculation gives
\bea
&&N_{1} = 4m^2 \left\{ \left(k_a -k_b \right)_{0} \ub \ks_a u + 2 
\left(k_b - k_c \right)_0 \ub \ks_c u + \right. \\ \nn
&&\left. + \ub \left(\ks_b -\ks_c \right) \ks_a u + \ub \ks_a \left(
\ks_c - \ks_a \right) u 
- \ub  \ks_c \left( \ks_a - \ks_b \right) u - 
\ub \left( \ks_c -\ks_a \right) \ks_c u \right\} 
=N_{1c}
\eea
where the rest-frame condition was used. An entirely similar calculation gives 
$N_2 = N_{2d}$ to linear accuracy. 
Thus the numerator factors are
equal up to linear order in the integration, and equality
(\ref{eq:ward25}) is established to linear order. Hence we have shown 
the cancellation
of the logarithmic and linear wave function renormalization pieces of group-5.



\begin{figure}[p]
\begin{center} \begin{picture}(125,80)(0,0)

\SetScale{1}
\SetWidth{0.5}
\ArrowLine(7,7)(21,70)
\ArrowLine(105,70)(119,7)
\SetScale{7}
\SetWidth{0.071}
\Gluon(9,5)(9,10){-0.23}{7}
\Gluon(16.5,3.25)(9,5){-0.23}{11}
\Gluon(1.7,3.25)(9,5){0.23}{11}
\SetWidth{0.01}
\GOval(9,10)(1,6)(0){0.7}
\SetWidth{0.071}
\DashCArc(9,5)(3,-13.134,90){0.2}
\DashCArc(9,5)(3,90,193.134){0.2}
\BCirc(9,5){1.5}
\Text(63,35)[]{S}

\end{picture}

\end{center}
\caption{\setlength{\baselineskip}{0.30in} 
\setlength{\baselineskip}{0.30in}
The forward scattering amlitude for the heavy quark decay whose 
imaginary part gives the decay width. Shaded blob denotes
the hard part, while S denotes the soft gluon interactions.}
\label{fig:fsa}
\end{figure}
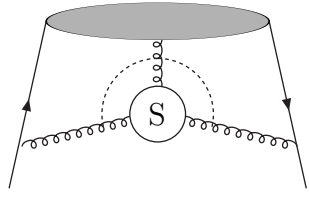


\begin{figure}
\begin{center} \begin{picture}(400,90)(0,0)

\SetOffset(0,12)
\SetScale{7}
\SetWidth{0.071}

\Line(1,1)(3,10)
\Line(17,1)(15,10)
\Line(3,10)(15,10)
\Gluon(9,5)(9,10){-0.23}{7}
\SetWidth{0.01}
\GOval(9,5)(1,7)(0){0.7}
\SetWidth{0.071}
\Text(63,0)[]{a)}
\Text(42,77)[]{$\large$ $p_1-k$}
\Text(84,77)[]{$\large$ $p_1$}
\Text(70,56)[]{$\large$ $k$}

\SetOffset(135,12)
\Line(1,1)(3,10)
\Line(17,1)(15,10)
\Line(3,10)(15,10)
\Gluon(9,5)(9,10){-0.23}{7}
\Gluon(16.5,3.25)(9,5){-0.23}{11}
\Gluon(1.5,3.25)(9,5){0.23}{11}
\SetWidth{0.071}
\Text(63,0)[]{b)}

\SetOffset(270,12)
\Line(1,1)(3,10)
\Line(17,1)(15,10)
\Line(3,10)(15,10)
\Gluon(6,5)(6,10){-0.23}{7}
\Gluon(12,5)(12,10){-0.23}{7}
\SetWidth{0.01}
\GOval(9,5)(1,7)(0){0.7}
\SetWidth{0.071}
\Text(63,0)[]{c)}

\end{picture}

\end{center}
\caption{\setlength{\baselineskip}{0.30in}
a) Emission of a single gluon from the final state quark.
The blob denotes all possible gluon self-interactions before
coupling to the heavy quark. b) An example for the gluon self-interactions
included in a). c) Two gluon emission from the final state quark.}
\label{fig:cut}
\end{figure}
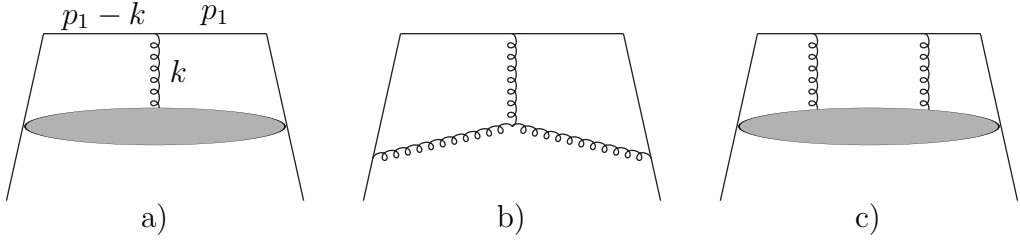


\begin{figure}
\begin{center} \begin{picture}(400,90)(0,0)

\SetOffset(0,10)
\SetScale{7}
\SetWidth{0.071}

\SetScale{1}
\SetWidth{0.5}
\ArrowLine(7,7)(35,70)
\SetScale{7}
\SetWidth{0.071}
\Text(35,0)[]{a)}

\Line(9,1)(5,10)
\Vertex(5,10){0.5}

\Text(11,40)[]{$\large$ $p$}
\Text(80,35)[]{$\large$ $\ps^5$}

\end{picture}

\end{center}

\begin{center} \begin{picture}(400,100)(0,0)

\SetScale{7}
\SetWidth{0.071}

\SetScale{1}
\SetWidth{0.5}
\ArrowLine(7,28)(35,91)
\SetScale{7}
\SetWidth{0.071}
\Line(9,4)(5,13)
\Gluon(5,6)(5,13){-0.23}{9}
\BCirc(5,13){0.7}
\Line(5.5,12.5)(4.5,13.5) \Line(4.5,12.5)(5.5,13.5)

\Text(35,9)[]{b)}
\Text(11,61)[]{$\large$ $p$}
\Text(35,36)[c]{$\large$ $\uparrow$}
\Text(35,25)[]{$\large$ $k, \alpha, a$}

\Text(60,61)[l]{$\large$ $g~T^a~[(\ps+\ks)^4 \gamma_{\alpha} +
(\ps+\ks)^3 \gamma_{\alpha}
\ps+ (\ps+\ks)^2 \gamma_{\alpha} \ps^2 + (\ps+\ks) 
\gamma_{\alpha} \ps^3 +
\gamma_{\alpha} \ps^4]$}

\end{picture}

\end{center}

\begin{center} \begin{picture}(400,100)(0,0)

\SetScale{7}
\SetWidth{0.071}

\SetOffset(0,16)

\SetScale{1}
\SetWidth{0.5}
\ArrowLine(7,14)(35,77)
\SetScale{7}
\SetWidth{0.071}
\Text(11,47)[]{$\large$ $p$}
\Text(35,-7)[]{c)}

\Line(9,2)(5,11)
\Gluon(4,4)(5,11){-0.23}{9}
\Gluon(6,4)(5,11){0.23}{9}
\BCirc(5,11){0.7}
\Line(5.5,10.5)(4.5,11.5) \Line(4.5,10.5)(5.5,11.5)

\Text(26,21)[]{$\large$ $\uparrow$}
\Text(40,21)[]{$\large$ $\uparrow$}
\Text(21,16)[]{$k_1$}
\Text(21,5)[]{$\alpha, a$}
\Text(50,16)[]{$k_2$}
\Text(50,5)[]{$\beta, b$}
 
\Text(70,60)[l]{$\large$
$g^2~\left[ \left(\ps+ \ks_1+ \ks_2 \right)^3
\left(\gamma_{\alpha} \gamma_{\beta} T^a T^b + \gamma_{\beta}
\gamma_{\alpha}T^b T^a \right)\right.$}
\Text(70,41)[l]{$\large$
$+~(\ps+\ks_1+\ks_2)^2 \left(\gamma_{\beta} T^b \left(\ps+\ks_1 \right)
\gamma_{\alpha} T^a + \gamma_{\alpha} T^a \left(\ps+\ks_2 \right)
\gamma_{\beta} T^b \right)$}
\Text(70,23)[l]{$\large$
$\left. +~\cdots~+\left(\gamma_{\alpha} \gamma_{\beta} T^a T^b
+ \gamma_{\beta}\gamma_{\alpha}T^b T^a \right) \ps^3 \right]$}

\end{picture}

\end{center}

\caption{\setlength{\baselineskip}{0.30in}
a) leading effective vertex,  
b) single gluon effective vertex ,
c) double gluon effective vertex}
\label{fig:effvertex}

\end{figure}
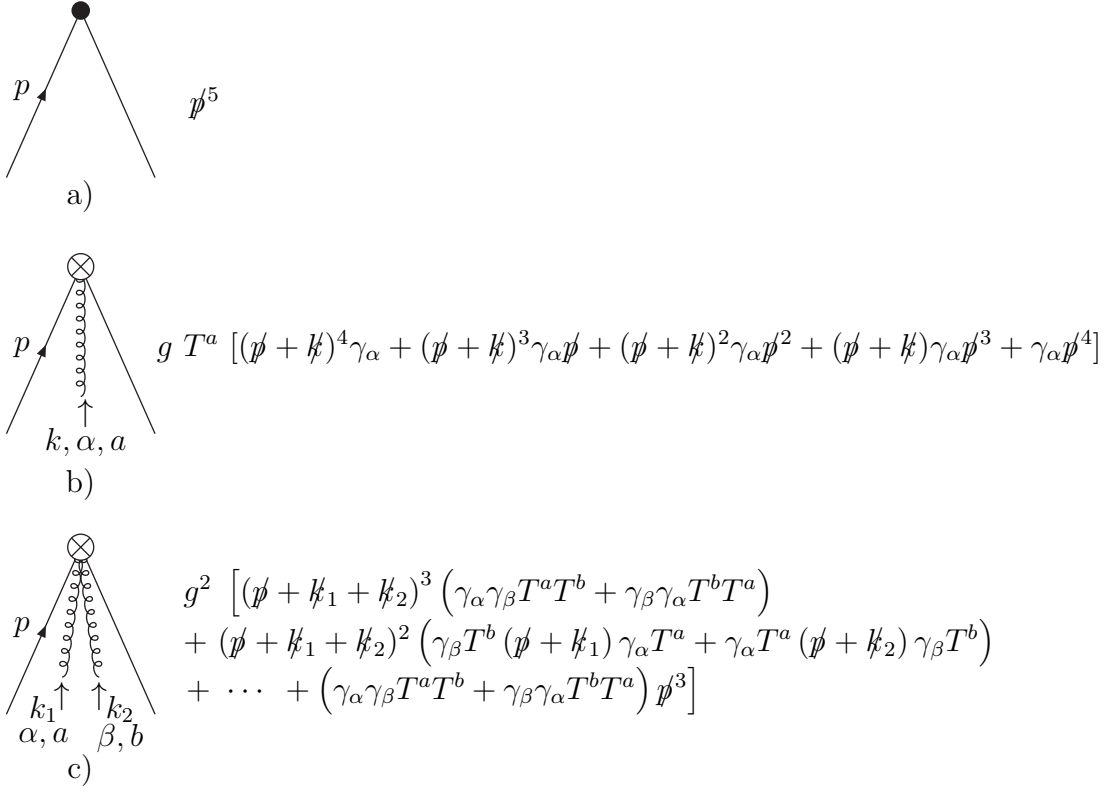


\begin{figure}
\begin{center} \begin{picture}(240,90)(0,0)

\SetScale{7}
\SetWidth{0.071}

\SetOffset(0,12)
\Line(1,1)(5,10)
\Line(9,1)(5,10)
\Vertex(5,10){0.5}
\GlueArc(3,5.5)(1.7,66,246){-0.23}{9}
\Text(35,0)[]{a)}

\SetOffset(84,12)
\Line(1,1)(5,10)
\Line(9,1)(5,10)
\Vertex(5,10){0.5}
\GlueArc(7,5.5)(1.7,-66,114){-0.23}{9}
\Text(35,0)[]{b)}

\SetOffset(168,12)
\Line(1,1)(5,10)
\Line(9,1)(5,10)
\Vertex(5,10){0.5}
\Gluon(2.55,4.5)(7.45,4.5){-0.23}{9}
\Text(35,0)[]{c)}

\end{picture}

\end{center}
\caption{\setlength{\baselineskip}{0.30in}
Wave function renormalization and Bremsstrahlung diagrams 
to order $g^2$.} 
\label{fig:wave1}

\end{figure}


\begin{figure}
\begin{center} \begin{picture}(210,63)(0,0)

\SetScale{9}
\SetWidth{0.045}

\Text(20,32)[]{$-i$\Large$\frac{\partial}{\partial p_{\nu}}$}

\SetOffset(32,24)
\Line(1,1)(9,1)
\GlueArc(5,1)(1.7,0,180){-0.23}{9}

\SetOffset(0,0)
\Text(122,32)[]{\large $=$}

\SetOffset(120,24)
\Line(1,1)(9,1)
\GlueArc(5,1)(1.7,0,180){-0.23}{9}
\DashLine(5,1)(5,-2){0.3}
\Text(51,-13)[]{$\nu$}

\end{picture}

\end{center}
\caption{\setlength{\baselineskip}{0.30in}
One-loop order Ward-like identity}
\label{fig:ward1}

\end{figure}


\begin{figure}
\begin{center} \begin{picture}(350,90)(0,0)

\SetScale{7}
\SetWidth{0.071}

\SetOffset(0,12)
\Line(1,1)(5,10)
\Line(9,1)(5,10)
\Vertex(5,10){0.5}
\Gluon(2.55,4.5)(7.45,4.5){-0.23}{9}
\Text(43.57,50.75)[]{$/$}
\Text(35,0)[]{a)}

\Text(75,40)[]{\large $+$}
\SetOffset(84,12)
\Line(1,1)(5,10)
\Line(9,1)(5,10)
\GlueArc(4.31,8.45)(1.7,66,246){-0.23}{9}
\BCirc(5,10){0.7}
\Line(5.5,9.5)(4.5,10.5) \Line(4.5,9.5)(5.5,10.5)
\Text(35,0)[]{b)}

\Text(75,40)[]{\large $+$}

\SetOffset(168,12)
\Line(1,1)(5,10)
\Line(9,1)(5,10)
\GlueArc(5.7,8.45)(1.7,-66,114){-0.23}{9}
\BCirc(5,10){0.7}
\Line(5.5,9.5)(4.5,10.5) \Line(4.5,9.5)(5.5,10.5)
\Text(35,0)[]{c)}

\Text(75,40)[]{\large $+$}

\SetOffset(252,12)
\Line(1,1)(5,10)
\Line(9,1)(5,10)
\Vertex(5,10){0.5}
\Text(71,7)[]{\large $\Sigma_1$}
\Text(35,0)[]{d)}

\Text(75,40)[]{\large $=$}
\Text(92,40)[]{\large $0$}
\Text(75,49)[]{\small linear}

\end{picture}

\end{center}

\caption{\setlength{\baselineskip}{0.30in}
Mass renormalization and the corresponding bremsstrahlung diagram to 
order $g^2$.
Figure a) depicts the mass renormalization piece of the graph. The
propagator with slash is cancelled by the expansion of the 
leading effective vertex.
The subscript $\Sigma_1$ of d) denotes the mass
shift due to the order $g^2$ self-energy correction.} 
\label{fig:mass1}
\end{figure}


\begin{figure}
\begin{center} \begin{picture}(380,63)(0,0)

\SetScale{9}
\SetWidth{0.045}

\Text(20,32)[]{$-i$\Large$ \Sigma =$}

\SetOffset(32,24)
\Line(1,1)(9,1)
\GlueArc(5,1)(1.7,0,180){-0.23}{9}

\Text(40,-10)[]{$-i$\large$ \Sigma_1$}
\SetOffset(0,0)
\Text(123,32)[]{\large $+$}

\SetOffset(124,24)
\Line(1,1)(9,1)
\GlueArc(5,1)(1.7,0,180){-0.23}{9}
\GlueArc(5,1)(0.8,0,180){-0.23}{4}

\Text(40,-10)[]{$-i$\large$ \Sigma_{21}$}

\SetOffset(0,0)
\Text(215,32)[]{\large $+$}

\SetOffset(216,24)
\Line(1,1)(9,1)
\GlueArc(4.15,1)(1.7,0,180){-0.23}{9}
\GlueArc(5.85,1)(1.7,-180,0){-0.23}{9}

\Text(32,-10)[]{$-i$\large$ \Sigma_{22}$}

\SetOffset(0,0)
\Text(304,32)[]{\large $+$}

\SetOffset(302,24)
\Line(1,1)(9,1)
\GlueArc(5,1)(1.7,0,180){-0.23}{9}
\Text(45,9)[]{\large$\star$}

\Text(40,-10)[]{$-i$\large$ \Sigma_{23}$}
\end{picture}
\end{center}

\begin{center} \begin{picture}(380,63)(0,0)

\SetScale{9}
\SetWidth{0.045}

\Text(33,32)[]{\large $+$}

\SetOffset(32,24)
\Line(1,1)(9,1)
\GlueArc(5,1)(1.7,0,180){-0.23}{9}
\GCirc(5,2.7){0.5}{0.45}

\Text(40,-10)[]{$-i$\large$ \Sigma_{24}$}

\SetOffset(0,0)
\Text(123,32)[]{\large $+$}

\SetOffset(124,24)
\Line(1,1)(9,1)
\GlueArc(5,1)(1.7,0,180){-0.23}{9}
\Gluon(5,1)(5,2.7){0.23}{3}

\Text(40,-10)[]{$-i$\large$ \Sigma_{25}$}

\end{picture}
\end{center}
\caption{\setlength{\baselineskip}{0.30in}
Self-energy corrections of order $g^2$ and $g^4$}  
\label{fig:selfen}
\end{figure}


\begin{figure}
\begin{center} \begin{picture}(400,90)(0,0)

\SetScale{7}
\SetWidth{0.071}

\SetOffset(0,12)
\Line(1,1)(5,10)
\Line(9,1)(5,10)
\Vertex(5,10){0.5}
\GlueArc(3,5.5)(1.7,66,246){-0.23}{9}
\GlueArc(3,5.5)(0.8,66,246){-0.23}{4}
\Text(35,0)[]{a)}

\SetOffset(84,12)
\Line(1,1)(5,10)
\Vertex(5,10){0.5}
\Line(9,1)(5,10)
\GlueArc(7,5.5)(1.7,-66,114){-0.23}{9}
\GlueArc(7,5.5)(0.8,-66,114){-0.23}{4}
\Text(35,0)[]{b)}

\SetOffset(168,12)
\Line(1,1)(5,10)
\Line(9,1)(5,10)
\Vertex(5,10){0.5}
\Gluon(1.88,3)(8.12,3){-0.23}{9}
\GlueArc(3.22,6)(1.7,66,246){-0.23}{9}
\Text(35,0)[]{c)}

\SetOffset(252,12)
\Line(1,1)(5,10)
\Line(9,1)(5,10)
\Vertex(5,10){0.5}
\Gluon(1.88,3)(8.12,3){-0.23}{9}
\Gluon(2.55,4.5)(7.45,4.5){-0.23}{7}
\Text(35,0)[]{d)}

\SetOffset(336,12)
\Line(1,1)(5,10)
\Line(9,1)(5,10)
\Text(35,0)[]{c)}
\Text(35,0)[]{c)}
\Vertex(5,10){0.5}
\Gluon(1.88,3)(8.12,3){-0.23}{9}
\GlueArc(6.78,6)(1.7,-66,114){-0.23}{9}
\Text(35,0)[]{e)}

\end{picture}

\end{center}
\caption{\setlength{\baselineskip}{0.30in}
The first group of 2-loop diagrams corresponding to $\Sigma_{21}$.}
\label{fig:wave21}

\end{figure}
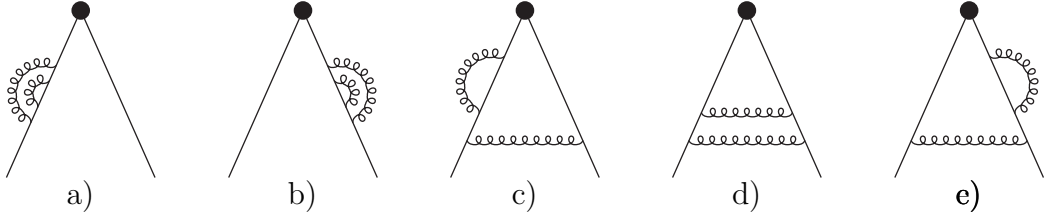


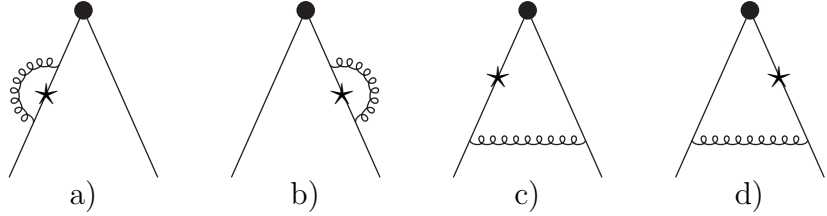
\begin{figure}
\begin{center} \begin{picture}(320,90)(0,0)

\SetScale{7}
\SetWidth{0.071}

\SetOffset(0,12)
\Line(1,1)(5,10)
\Line(9,1)(5,10)
\Vertex(5,10){0.5}
\Text(21,38.5)[]{\Large $\star$}
\GlueArc(3,5.5)(1.7,66,246){-0.23}{9}
\Text(35,0)[]{a)}

\SetOffset(84,12)
\Line(1,1)(5,10)
\Line(9,1)(5,10)
\Vertex(5,10){0.5}
\GlueArc(7,5.5)(1.7,-66,114){-0.23}{9}
\Text(49,38.5)[]{\Large $\star$}
\Text(35,0)[]{b)}

\SetOffset(168,12)
\Line(1,1)(5,10)
\Line(9,1)(5,10)
\Vertex(5,10){0.5}
\Text(23.8469,44.8966)[]{\Large $\star$}
\Gluon(1.88,3)(8.12,3){-0.23}{9}
\Text(35,0)[]{c)}

\SetOffset(252,12)
\Line(1,1)(5,10)
\Line(9,1)(5,10)
\Vertex(5,10){0.5}
\Text(46.1531,44.8966)[]{\Large $\star$}
\Gluon(1.88,3)(8.12,3){-0.23}{9}
\Text(35,0)[]{d)}

\end{picture}

\end{center}
\caption{\setlength{\baselineskip}{0.30in}
The third group of 2-loop diagrams corresponding to $\Sigma_{23}$.}
\label{fig:wave23}

\end{figure}


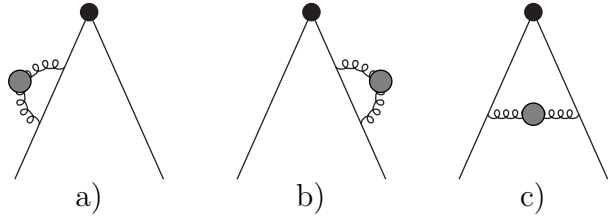
\begin{figure}
\begin{center} \begin{picture}(240,90)(0,0)

\SetScale{7}
\SetWidth{0.071}

\SetOffset(0,12)
\Line(1,1)(5,10)
\Line(9,1)(5,10)
\Vertex(5,10){0.5}
\GlueArc(3,5.5)(1.7,66,246){-0.23}{9}
\GCirc(1.27,6.27){0.6}{0.5}
\Text(35,0)[]{a)}

\SetOffset(84,12)
\Line(1,1)(5,10)
\Line(9,1)(5,10)
\Vertex(5,10){0.5}
\GlueArc(7,5.5)(1.7,-66,114){-0.23}{9}
\GCirc(8.73,6.27){0.6}{0.5}
\Text(35,0)[]{b)}

\SetOffset(168,12)
\Line(1,1)(5,10)
\Line(9,1)(5,10)
\Vertex(5,10){0.5}
\Gluon(2.55,4.5)(7.45,4.5){-0.23}{9}
\GCirc(5,4.5){0.6}{0.5}
\Text(35,0)[]{c)}

\end{picture}

\end{center}

\caption{\setlength{\baselineskip}{0.30in}
The 4th group of 2-loop diagrams corresponding to $\Sigma_{24}$.
The blob denotes fermion, gluon and ghost loops.}
\label{fig:wave24}

\end{figure}

\begin{figure}
\begin{center} \begin{picture}(320,90)(0,0)

\SetScale{7}
\SetWidth{0.071}

\SetOffset(0,12)
\Line(1,1)(5,10)
\Line(9,1)(5,10)
\Vertex(5,10){0.5}
\GlueArc(3,5.5)(1.7,66,246){-0.23}{9}
\Gluon(3,5.5)(1.27,6.27){-0.23}{3}
\Text(35,0)[]{a)}

\SetOffset(84,12)
\Line(1,1)(5,10)
\Line(9,1)(5,10)
\Vertex(5,10){0.5}
\GlueArc(7,5.5)(1.7,-66,114){-0.23}{9}
\Gluon(7,5.5)(8.73,6.27){-0.23}{3}
\Text(35,0)[]{b)}

\SetOffset(168,12)
\Line(1,1)(5,10)
\Line(9,1)(5,10)
\Vertex(5,10){0.5}
\Gluon(3.88,7.5)(7.88,3.5){-0.23}{9}
\Gluon(5,6.0)(2.12,3.5){0.23}{8}
\Text(35,0)[]{c)}

\SetOffset(252,12)
\Line(1,1)(5,10)
\Line(9,1)(5,10)
\Vertex(5,10){0.5}
\Gluon(2.12,3.5)(6.12,7.5){-0.23}{9}
\Gluon(5,6)(7.88,3.5){-0.23}{8}
\Text(35,0)[]{d)}

\end{picture}

\end{center}

\caption{\setlength{\baselineskip}{0.30in}
The 5th group of 2-loop diagrams corresponding to $\Sigma_{25}$.}
\label{fig:wave25}

\end{figure}
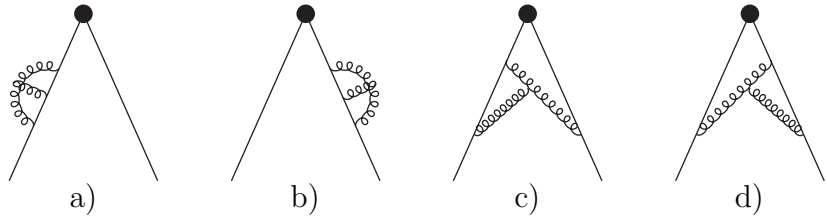


\begin{figure}
\begin{center} \begin{picture}(380,63)(0,0)

\SetScale{9}
\SetWidth{0.045}

\Text(20,32)[]{$-i$\Large$ \frac{\partial}{\partial p_{\nu}}$}

\SetOffset(32,24)
\Line(1,1)(9,1)
\GlueArc(5,1)(1.7,0,180){-0.23}{9}
\GlueArc(5,1)(0.8,0,180){-0.23}{4}

\SetOffset(0,0)
\Text(123,32)[]{\large $=$}

\SetOffset(124,24)
\Line(1,1)(9,1)
\GlueArc(5,1)(1.7,0,180){-0.23}{9}
\GlueArc(5,1)(0.8,0,180){-0.23}{4}

\DashLine(3.7,1)(3.7,-2){0.3}
\Text(39,-13)[]{$\nu$}

\SetOffset(0,0)
\Text(215,32)[]{\large $+$}
\Text(215,1)[]{a)}

\SetOffset(216,24)
\Line(1,1)(9,1)
\GlueArc(5,1)(1.7,0,180){-0.23}{9}
\GlueArc(5,1)(0.8,0,180){-0.23}{4}

\DashLine(5,1)(5,-2){0.3}
\Text(51,-13)[]{$\nu$}

\SetOffset(0,0)
\Text(304,32)[]{\large $+$}

\SetOffset(302,24)
\Line(1,1)(9,1)
\GlueArc(5,1)(1.7,0,180){-0.23}{9}
\GlueArc(5,1)(0.8,0,180){-0.23}{4}

\DashLine(6.3,1)(6.3,-2){0.3}
\Text(62,-13)[]{$\nu$}
\end{picture}
\end{center}

\begin{center} \begin{picture}(380,63)(0,0)

\SetScale{9}
\SetWidth{0.045}

\Text(20,32)[]{$-i$\Large$ \frac{\partial}{\partial p_{\nu}}$}

\SetOffset(32,24)
\Line(1,1)(9,1)
\GlueArc(4.15,1)(1.7,0,180){-0.23}{9}
\GlueArc(5.85,1)(1.7,-180,0){-0.23}{9}

\SetOffset(0,0)
\Text(123,32)[]{\large $=$}

\SetOffset(124,24)
\Line(1,1)(9,1)
\GlueArc(4.15,1)(1.7,0,180){-0.23}{9}
\GlueArc(5.85,1)(1.7,-180,0){-0.23}{9}

\DashLine(3.3,1)(3.3,-2){0.3}
\Text(36,-13)[]{$\nu$}

\SetOffset(0,0)
\Text(215,32)[]{\large $+$}
\Text(215,7)[]{b)}

\SetOffset(216,24)
\Line(1,1)(9,1)
\GlueArc(4.15,1)(1.7,0,180){-0.23}{9}
\GlueArc(5.85,1)(1.7,-180,0){-0.23}{9}

\DashLine(5,1)(5,-2){0.3}
\Text(51,-13)[]{$\nu$}

\SetOffset(0,0)
\Text(304,32)[]{\large $+$}

\SetOffset(302,24)
\Line(1,1)(9,1)
\GlueArc(4.15,1)(1.7,0,180){-0.23}{9}
\GlueArc(5.85,1)(1.7,-180,0){-0.23}{9}

\DashLine(6.7,1)(6.7,-2){0.3}
\Text(66.3,-13)[]{$\nu$}
\end{picture}
\end{center}

\begin{center} \begin{picture}(380,63)(0,0)

\SetScale{9}
\SetWidth{0.045}

\Text(20,32)[]{$-i$\Large$ \frac{\partial}{\partial p_{\nu}}$}

\SetOffset(32,24)
\Line(1,1)(9,1)
\GlueArc(5,1)(1.7,0,180){-0.23}{9}
\Text(45,9)[]{\Large$\star$}

\SetOffset(0,0)
\Text(123,32)[]{\large $=$}

\SetOffset(124,24)
\Line(1,1)(9,1)
\GlueArc(5,1)(1.7,0,180){-0.23}{9}
\Text(45,9)[]{\Large$\star$}

\DashLine(4.15,1)(4.15,-2){0.3}
\Text(43.35,-13)[]{$\nu$}

\SetOffset(0,0)
\Text(215,32)[]{\large $+$}
\Text(215,1)[]{c)}

\SetOffset(216,24)
\Line(1,1)(9,1)
\GlueArc(5,1)(1.7,0,180){-0.23}{9}
\Text(45,9)[]{\Large$\star$}

\DashLine(5.85,1)(5.85,-2){0.3}
\Text(58.65,-13)[]{$\nu$}

\end{picture}

\end{center}

\begin{center} \begin{picture}(380,63)(0,0)

\SetScale{9}
\SetWidth{0.045}

\Text(20,32)[]{$-i$\Large$ \frac{\partial}{\partial p_{\nu}}$}

\SetOffset(32,24)
\Line(1,1)(9,1)
\GlueArc(5,1)(1.7,0,180){-0.23}{9}
\GCirc(5,2.7){0.5}{0.45}

\SetOffset(0,0)
\Text(123,32)[]{\large $=$}

\SetOffset(124,24)
\Line(1,1)(9,1)
\GlueArc(5,1)(1.7,0,180){-0.23}{9}
\GCirc(5,2.7){0.5}{0.45}

\DashLine(5,1)(5,-2){0.3}
\Text(51,-13)[]{$\nu$}
\SetOffset(0,0)
\Text(215,1)[]{d)}

\end{picture}
\end{center}

\begin{center} \begin{picture}(380,63)(0,0)

\SetScale{9}
\SetWidth{0.045}

\Text(20,32)[]{$-i$\Large$ \frac{\partial}{\partial p_{\nu}}$}

\SetOffset(32,24)
\Line(1,1)(9,1)
\GlueArc(5,1)(1.7,0,180){-0.23}{9}
\Gluon(5,1)(5,2.7){0.23}{3}

\SetOffset(0,0)
\Text(123,32)[]{\large $=$}

\SetOffset(124,24)
\Line(1,1)(9,1)
\GlueArc(5,1)(1.7,0,180){-0.23}{9}
\Gluon(5,1)(5,2.7){0.23}{3}

\DashLine(4.15,1)(4.15,-2){0.3}
\Text(43.35,-13)[]{$\nu$}

\SetOffset(0,0)
\Text(215,32)[]{\large $+$}
\Text(215,1)[]{e)}

\SetOffset(216,24)
\Line(1,1)(9,1)
\GlueArc(5,1)(1.7,0,180){-0.23}{9}
\Gluon(5,1)(5,2.7){0.23}{3}

\DashLine(5.85,1)(5.85,-2){0.3}
\Text(58.65,-13)[]{$\nu$}

\end{picture}

\end{center}
\caption{\setlength{\baselineskip}{0.30in}
Two-loop order Ward-like identities}
\label{fig:ward2}

\end{figure}


\begin{figure}
\begin{center} \begin{picture}(300,90)(0,0)

\SetScale{7}
\SetWidth{0.071}

\SetOffset(8,0)
\Line(1,1)(5,10)
\Line(9,1)(5,10)
\GlueArc(1.8122,2.827)(1.5,66,246){-0.23}{7}
\GlueArc(3.8,7.39)(1.5,66,246){-0.23}{7}
\Vertex(5,10){0.5}

\Text(75,40)[]{\large $+$}

\SetOffset(84,0)
\Line(1,1)(5,10)
\Line(9,1)(5,10)
\Vertex(5,10){0.5}
\GlueArc(8.2,2.827)(1.5,-66,114){-0.23}{7}
\GlueArc(6.2,7.39)(1.5,-66,114){-0.23}{7}

\Text(75,40)[]{\large $+$}
\Text(90,40)[l]{\large $\cdots$}

\SetOffset(0,0)
\Text(200,41)[]{\large $=$}
\Text(270,41)[]{\large $\frac{3}{4} \left[ (Z^{-1}-1)_{g^2} \right]^2 m_p^5$}

\end{picture}

\end{center}

\begin{center} \begin{picture}(300,90)(0,0)

\SetScale{7}
\SetWidth{0.071}

\SetOffset(84,0)
\Line(1,1)(5,10)
\Line(9,1)(5,10)
\Vertex(5,10){0.5}
\GlueArc(3,5.5)(1.7,66,246){-0.23}{9}
\GlueArc(7,5.5)(1.7,-66,114){-0.23}{9}

\Text(75,40)[]{\large $+$}
\Text(90,40)[l]{\large $\cdots$}

\SetOffset(0,0)
\Text(200,41)[]{\large $=$}
\Text(270,41)[]{\large $\frac{1}{4} \left[ (Z^{-1}-1)_{g^2} \right]^2 m_p^5$}

\end{picture}

\end{center}

\begin{center} \begin{picture}(300,90)(0,0)

\SetScale{7}
\SetWidth{0.071}

\SetOffset(8,0)
\Line(1,1)(5,10)
\Line(9,1)(5,10)
\GlueArc(1.8122,2.827)(1.5,66,246){-0.23}{7}
\Gluon(3,5.55)(7,5.55){-0.23}{5}
\Vertex(5,10){0.5}

\Text(75,40)[]{\large $+$}

\SetOffset(84,0)
\Line(1,1)(5,10)
\Line(9,1)(5,10)
\Vertex(5,10){0.5}
\Gluon(3,5.55)(7,5.55){-0.23}{5}
\GlueArc(8.2,2.827)(1.5,-66,114){-0.23}{7}

\Text(75,40)[]{\large $+$}
\Text(90,40)[l]{\large $\cdots$}

\SetOffset(0,0)
\Text(200,41)[]{\large $=$}
\Text(270,41)[]{\large $-\left[ (Z^{-1}-1)_{g^2} \right]^2 m_p^5$}

\end{picture}

\end{center}
\caption{\setlength{\baselineskip}{0.30in}
1-particle reducible diagrams for wave function 
renormalization. 
The ellipsis denote all possible inclusion of mass counterterms.}
\label{fig:1pr}

\end{figure}


\begin{figure}
\begin{center} \begin{picture}(380,90)(0,0)

\SetScale{7}
\SetWidth{0.071}

\SetOffset(0,12)
\Line(1,1)(5,10)
\Line(9,1)(5,10)
\Vertex(5,10){0.5}
\Gluon(3.88,7.5)(7.88,3.5){-0.23}{9}
\Gluon(5,6.0)(2.12,3.5){0.23}{8}
\Text(75,40)[]{$\large$ $+$}
\Text(43.57,50.75)[]{$/$}
\Text(35,0)[]{a)}

\SetOffset(84,12)
\Line(1,1)(5,10)
\Line(9,1)(5,10)
\Gluon(3.86,7.44)(1.30,8.58){0.23}{7}
\GlueArc(3.86,7.44)(2.8,66,246){-0.23}{13}
\BCirc(5,10){0.7}
\Line(5.5,9.5)(4.5,10.5) \Line(4.5,9.5)(5.5,10.5)
\Text(35,0)[]{b)}

\Text(75,40)[]{\large $=$}
\Text(92,40)[]{\large $0$}
\Text(75,49)[]{\small linear}

\SetOffset(188,12)
\Line(1,1)(5,10)
\Line(9,1)(5,10)
\Vertex(5,10){0.5}
\Gluon(2.12,3.5)(6.12,7.5){-0.23}{9}
\Gluon(5,6)(7.88,3.5){-0.23}{8}
\Text(26.43,50.75)[]{$\backslash$}
\Text(35,0)[]{c)}

\Text(75,40)[]{$\large$ $+$}

\SetOffset(272,12)
\Line(1,1)(5,10)
\Line(9,1)(5,10)
\Gluon(6.14,7.44)(8.69,8.58){0.23}{7}
\GlueArc(6.14,7.44)(2.8,-66,114){-0.23}{13}
\BCirc(5,10){0.7}
\Line(5.5,9.5)(4.5,10.5) \Line(4.5,9.5)(5.5,10.5)
\Line(5.5,9.5)(4.5,10.5) \Line(4.5,9.5)(5.5,10.5)
\Text(35,0)[]{d)}

\Text(79,40)[]{\large $=$}
\Text(96,40)[]{\large $0$}
\Text(79,49)[]{\small linear}
\end{picture}
\end{center}

\begin{center} \begin{picture}(380,90)(0,0)

\SetScale{7}
\SetWidth{0.071}

\SetOffset(0,12)
\Line(1,1)(5,10)
\Line(9,1)(5,10)
\Vertex(5,10){0.5}
\Text(71,7)[]{\large $\Sigma_{25}$}
\Text(35,0)[]{e)}

\Text(75,40)[]{\large $+$}

\SetOffset(84,12)
\Line(1,1)(5,10)
\Line(9,1)(5,10)
\Gluon(5,5)(5,10){0.23}{8}
\Gluon(2,3)(5,5){0.23}{6}
\Gluon(5,5)(8,3){0.23}{6}
\BCirc(5,10){0.7}
\Line(5.5,9.5)(4.5,10.5) \Line(4.5,9.5)(5.5,10.5)
\Text(35,0)[]{f)}

\Text(75,40)[]{\large $+$}

\SetOffset(168,12)
\Line(1,1)(5,10)
\Line(9,1)(5,10)
\Gluon(2,3)(5,5){0.23}{6}
\GlueArc(10,7.5)(5.59,153.44,206.56){0.2}{8}
\GlueArc(0,7.5)(5.59,-26.56,26.56){0.2}{8}
\BCirc(5,10){0.7}
\Line(5.5,9.5)(4.5,10.5) \Line(4.5,9.5)(5.5,10.5)
\Text(35,0)[]{g)}

\Text(75,40)[]{\large $+$}

\SetOffset(252,12)
\Line(1,1)(5,10)
\Line(9,1)(5,10)
\GlueArc(10,7.5)(5.59,153.44,206.56){0.2}{8}
\GlueArc(0,7.5)(5.59,-26.56,26.56){0.2}{8}
\Gluon(5,5)(8,3){0.23}{6}
\BCirc(5,10){0.7}
\Line(5.5,9.5)(4.5,10.5) \Line(4.5,9.5)(5.5,10.5)
\Text(35,0)[]{h)}

\Text(75,40)[]{\large $=$}
\Text(92,40)[]{\large $0$}
\Text(75,49)[]{\small linear}
\end{picture}

\end{center}
\caption{\setlength{\baselineskip}{0.30in}
Mass renormalization of order $g^4$: cancellation of linear terms 
in group-5.}
\label{fig:mass25} 
\end{figure}


\begin{thebibliography}{99}
\bibitem{fnote1} The wave function renormalization of the initial heavy quark is
another source of infrared sensitivity at the logarithmic and at the linear
level. These also cancel against the radiated energy as we discuss in detail in
the following sections, however for the purposes of the subsequent discussion, we
will consider just the mass renormalization contribution.
\end{thebibliography}
\end{document}